\documentclass[aps,prb,showpacs,twocolumn,floats,epsfig,pdflatex, superscriptaddress,longbibliography]{revtex4-2}
\usepackage{amssymb}
\usepackage{amsbsy}
\usepackage{amsmath}
\usepackage{epsfig}
\usepackage{amsfonts}
\usepackage{graphicx}
\usepackage{amssymb}
\usepackage{upgreek}
\usepackage{multirow}
\usepackage{dsfont}
\usepackage{float}
\usepackage{bm}
\usepackage{xcolor, comment}
\usepackage{hyperref}
\usepackage{tikz}
\begin{document}

\title{Tight-binding and density-functional study of the Raman tensor in two-dimensional massive Dirac fermion systems}
\author{Sel\c{c}uk Parlak}
\affiliation{D\'epartement de Physique, Institut Quantique and Regroupement Qu\'eb\'ecois sur les Mat\'eriaux de Pointe, Universit\'e de Sherbrooke, Sherbrooke, Qu\'ebec, Canada J1K 2R1}
\author{Abhishek Kumar}
\affiliation{D\'epartement de Physique, Institut Quantique and Regroupement Qu\'eb\'ecois sur les Mat\'eriaux de Pointe, Universit\'e de Sherbrooke, Sherbrooke, Qu\'ebec, Canada J1K 2R1}
\affiliation{Department of physics, Indian Institute of Technology Jammu, Jammu-181221, Jammu and Kashmir, India}
\author{Runhan Li}
\affiliation{D\'epartement de Physique, Institut Quantique and Regroupement Qu\'eb\'ecois sur les Mat\'eriaux de Pointe, Universit\'e de Sherbrooke, Sherbrooke, Qu\'ebec, Canada J1K 2R1}
\author{Maia G. Vergniory}
\affiliation{D\'epartement de Physique, Institut Quantique and Regroupement Qu\'eb\'ecois sur les Mat\'eriaux de Pointe, Universit\'e de Sherbrooke, Sherbrooke, Qu\'ebec, Canada J1K 2R1}
\affiliation{Donostia International Physics Center, Manuel de Lardizabal pasealekua 4, 20018 Donostia, Basque Country, Spain.}
\author{Ion Garate}
\affiliation{D\'epartement de Physique, Institut Quantique and Regroupement Qu\'eb\'ecois sur les Mat\'eriaux de Pointe, Universit\'e de Sherbrooke, Sherbrooke, Qu\'ebec, Canada J1K 2R1}
\date{\today}
\begin{abstract}
Recently, two unusual features were theoretically predicted for the Raman response of out-of-plane phonons in magnetic two-dimensional materials hosting massive Dirac fermions. First, the phase difference between certain Raman tensor elements was found to be quantized to $\pm \pi/2$, sensitive only to the sign of the Dirac fermion mass.  Second, a selection rule was identified in the Raman intensity under circularly polarized light, which generalizes the well-known optical valley selection rule.  These predictions were based on a low-energy effective model in the continuum approximation. Here, we test the robustness of those results for more realistic theoretical approaches. First, we calculate the Raman tensor for an electronic tight-binding model on a honeycomb lattice with broken time-reversal and inversion symmetries. Second, we compute the Raman tensor from density-functional theory for a monolayer of ferromagnetic 2H-RuCl$_2$. Both calculations corroborate the analytical results found in the continuum model, thereby theoretically confirming the peculiar behavior of the Raman tensor for two dimensional massive Dirac fermion systems.
\end{abstract}
\maketitle
\section{Introduction}
After decades of fruitful investigations \cite{rev4, ren2016topological,rev3, rev2, rev1}, two-dimensional (2D) Dirac materials remain in vogue due to their captivating electronic properties and potentially impactful applications.
The elemental building block capturing the low-energy electronic properties of these materials is the Hamiltonian of a single Dirac fermion,
\begin{equation}
\label{eq:h0}
h_0({\bf k}) = v_x k_x \sigma_x+v_y k_y \sigma_y+m\sigma_z,
\end{equation}
where $v_x$ and $v_y$ are the Dirac velocities, $k_x$ and $k_y$ are the components of the electronic wave vector (measured from the particular point in the Brillouin zone where the Dirac fermion is located) in the plane of the material, $\sigma_i$ are Pauli matrices denoting a pseudospin $1/2$, and $m$ is the Dirac mass.
The energy spectrum of  Eq. (\ref{eq:h0}) is comprised by a conduction and a valence band, each of them nondegenerate, separated at $k=0$ by an energy gap equal to $2|m|$.

While the energy bands of Eq. (\ref{eq:h0}) depend only on the moduli of the Dirac mass and the Dirac velocities, the electronic eigenstates are sensitive to their signs.
This sensitivity manifests most remarkably in the band-geometric properties of Eq. (\ref{eq:h0}).
For example, the Berry curvature for the valence (conduction) band is given by
\begin{equation}
\boldsymbol{\Omega}({\bf k})=(-)\hat{\bf z}\frac{m v_x v_y}{2(v_x^2 k_x^2+v_y^2 k_y^2+m^2)^{3/2}},
\end{equation}
while the Chern number of the same band reads
\begin{equation}
\label{eq:chern}
C=(-)\frac{1}{2}{\rm sgn}(v_x v_y m).
\end{equation}
Both quantities are sensitive to the signs of the Dirac mass and velocities.
In particular, the topological invariant $C$ only depends on the signs. 

The aforementioned sign dependences manifest strikingly in the electromagnetic response of the 2D Dirac fermion.
For example, in the DC response, Eq. (\ref{eq:h0}) leads to a half-quantized Hall effect \cite{bernevig2013topological} at zero temperature when the Fermi level is inside the gap (this will be the situation considered throughout the present work). The Hall voltage reverses sign when the sign of $v_x v_y m$ is switched. 
Concerning the AC electromagnetic response, right-circularly polarized light of frequency equal to the energy gap $2 |m|$ will or will not be absorbed by the 2D Dirac fermion depending on the sign of  $v_x v_y m$. If the sign is such that right-circularly polarized light is absorbed, then inverting the polarity switches off the absorption (or vice versa) \cite{wang2008valley, selection2, selection3}. This is commonly known as the optical valley selection rule. 

In Ref. \cite{selcuk:2024}, some of us wondered whether inelastic scattering of light, e.g. Raman scattering, might also display particular sensitivity to the sign of $v_x v_y m$.
In Raman spectroscopy, a polarized light with frequency $\omega_1$ and polarization ${\bf e}_1$ is irradiated onto the sample. 
This light couples to electrons in the material and excites a phonon mode $\lambda$ of frequency $\omega_0$  (the wave vectors of the light and the phonon will be neglected in the present work). 
The Raman efficiency \cite{loudon1963, Basko:2009}, defined as the ratio between (i)  the number of detected photons of frequency $\omega_2=\omega_1-\omega_0$ and polarization ${\bf e}_2$ per unit cross sectional area of the crystal per unit time, and (ii) the flux of incident photons, reads
\begin{equation}
\label{eq:rint}
 I_\lambda =\gamma \frac{\omega_2}{\omega_1} | \hat{\bf e}_1^\dagger\cdot {\bf R}_\lambda\cdot\hat{\bf e}_2|^2,
\end{equation}
where ${\bf R}_\lambda$ is the Raman tensor \cite{Cardona1982,loudon1963, Basko:2009} dependent on $\omega_1$ and $\omega_0$ as well as on the electronic band structure. The explicit form of the prefactor $\gamma$, which is independent of light frequencies, can be seen in Appendix \ref{sec:efficiency}. 
We will hereafter refer to $I_\lambda$ as the ``Raman intensity.'' 


In the backscattering configuration, where $\hat{\bf e}_{1}$ and $\hat{\bf e}_{2}$ are parallel to the $xy$-plane and photons propagate in the $z$ direction, the Raman tensor can effectively be expressed as 
\begin{equation}
    {\bf R}_\lambda=\begin{pmatrix}
    R_{xx} & R_{xy} \\
    R_{yx} & R_{yy}
    \end{pmatrix},
\end{equation}
where $R_{\alpha\beta}$ is a Raman tensor element with $\{\alpha,\beta\} \in \{x,y\}$. These elements are generally complex, $R_{\alpha\beta}=|R_{\alpha\beta}| \exp(i \phi_{\alpha\beta})$  \cite{Jin2019,Pimenta2021}.
The phases $\phi_{\alpha\beta}$ are not individually observable due to the modulus square factor in Eq.~(\ref{eq:rint}),   but the phase differences between different Raman tensor elements are measurable with a light that has nonzero angular momentum along the $z$ axis, see for example Eq.~(21) of Ref.~\cite{selcuk:2024}. 

Even though Raman scattering is a widely used tool to characterize the fundamental properties of 2D Dirac materials \cite{Wu2018,MALARD200951,Saito_2016,Cong2020,graziotto2024infrared}, analytical theories of the Raman tensor  have been scarce, if not nonexistent, with the exception of work on massless Dirac fermions in graphene \cite{Basko:2009} and recent work by some of us on massive Dirac fermions \cite{selcuk:2024}. 
In particular, prior to Ref. \cite{selcuk:2024}, it was not known whether the signs of the Dirac masses and velocities could influence the Raman response of the bulk 2D material. 
Reference \cite{selcuk:2024} started to fill this gap of knowledge by predicting that $R_{xy}=-R_{yx}$, $R_{yy}=R_{xx}$ and 
\begin{equation}
\label{eq:ratio}
\frac{R_{xy}}{R_{xx}}\simeq i\frac{\omega_1 v_y}{2 m v_x}
\end{equation}
for electrons described by Eq. (\ref{eq:h0}) and for a phonon mode involving out-of-plane atomic vibrations.
This expression was obtained under the conditions (i) $\omega_0\ll \omega_1$, which is commonly realized in experiment, and (ii) a low electronic decay rate.

The unexpected simplicity of Eq. (\ref{eq:ratio}) had two immediate consequences. 
First, in the case of circularly polarized incident and scattered light of frequency $\omega_1=2|m v_x/v_y|$, the Raman intensity became
\begin{equation}
\label{eq:sel}
I_\lambda\propto |R_{xx}|^2\left[1\pm {\rm sgn}(v_x v_y m)\right],
\end{equation}
where the upper (lower) sign applies when the incident and detected lights are left- (right-) circularly polarized. 
Thus, in the case of right-circularly polarized incident and scattered light of frequency $\omega_1=2|m v_x/v_y|$, there is an extinction of the Raman intensity when ${\rm sgn}(v_x v_y m)=1$.
If the incident and scattered lights are left-circularly polarized, the extinction takes place when ${\rm sgn}(v_x v_y m)=-1$.
This therefore gives way to a selection rule, which generalizes the valley selection rule found in the optical absorption of 2D massive Dirac fermion systems.

As a second direct consequence of Eq. (\ref{eq:ratio}), the phase difference between the Raman tensor elements $R_{xx}$ and $R_{xy}$ gave
\begin{equation}
    \label{eq:phasedif}
    \phi_{xy}-\phi_{xx}=\frac{\pi}{2} {\rm sgn}(v_x v_ym)
\end{equation}
and was therefore quantized. It is surprising that the phase difference should be independent of the magnitudes of $\omega_1$, the Dirac velocities and the Dirac mass. 
At first glance, it is tempting to relate Eqs. (\ref{eq:phasedif}) and (\ref{eq:chern}). 
Yet, further analysis in Ref. \cite{selcuk:2024} showed that the quantization of the phase difference is not associated to the electronic Chern number. 
Still, the result remains striking and not fully understood.

Reference~\cite{selcuk:2024} then considered models containing multiple copies of Eq.~(\ref{eq:h0}), to extrapolate the theory to 2D transition metal dichalcogenides. 
It was found that 2D Dirac materials with broken inversion and time-reversal symmetry would display properties consistent with Eqs. (\ref{eq:sel}) and (\ref{eq:phasedif}) when the frequency of the incident light was close to the energy gap of the Dirac material.

One natural criticism for the study of  Ref. \cite{selcuk:2024} is that it relied on simple low-energy effective models in the continuum approximation. 
It is licit to ask what happens to the peculiar selection rule and phase differences when we consider more sophisticated and realistic theoretical descriptions of the electronic structure. 
This is the question that we intend to address in the present work. 

We start in Sec.~\ref{sec:tight}, where we employ a tight-binding model of the honeycomb lattice with broken inversion and time-reversal symmetries. The Raman tensor is then calculated for this model, using a similar microscopic formalism as in Refs. \cite{Basko:2009, selcuk:2024}.
We further increase the degree of theoretical sophistication in Sec.~\ref{sec:first}, by performing a first-principles calculation on a monolayer of 2H-RuCl$_2$. This system has been recently predicted to be a ferromagnet with out-of-plane magnetization and low-energy Dirac fermions with an intrinsic valley Zeeman splitting \cite{runhan:2023}, making it an ideal candidate to test the theory of Ref. \cite{selcuk:2024} in a real material. 
In this section, the Raman tensor is calculated from density-functional theory (DFT), via the change of the electronic dielectric function under lattice vibrations. 
One advantage of this approach with respect to the one used in Sec.~\ref{sec:first} is that it does not require to compute electron-phonon matrix elements. One disadvantage of the method is that it neglects the phonon frequency.

In short, the results from Secs.~\ref{sec:tight} and \ref{sec:first} support the analytical theory of Ref. \cite{selcuk:2024} when $\omega_1$ is in the vicinity of an energy gap of one of the Dirac fermions. This conclusion is restated and the outlook for the project is presented in Sec.~\ref{sec:conc}.

\section{Tight-binding calculation}
\label{sec:tight}

In this section, we calculate the Raman tensor for a 2D tight-binding model of a honeycomb lattice with broken time- and space-inversion symmetries.

\subsection{Hamiltonian}

The calculation of the Raman tensor is based on the Hamiltonian 
\begin{equation}
    {\cal H}= {\cal H}_\text{e}+ {\cal H}_{\text{e-pt}}+ {\cal H}_{\text{e-pn}}+{\cal H}_{\rm e-pnpt},
\end{equation}
where ${\cal H}_\text{e}, {\cal H}_{\text{e-pt}}$ and ${\cal H}_{\text{e-pn}}$ are the electron, electron-photon and electron-phonon Hamiltonians, respectively, while ${\cal H}_{\rm e-pnpt}$ is the change in the electron-phonon Hamiltonian due to the presence of an electromagnetic field.
We now discuss each term of the Hamiltonian separately.
 
To begin with, the electronic  Hamiltonian is given by
\label{ham}
\begin{align}
{\cal H}_e &=\sum_{{\bf k}} \Psi_{\bf k}^\dag h_{\rm e}({\bf k}) \Psi_{\bf k},
\end{align}
 where ${\bf k}=(k_x, k_y)$ is the electronic wave vector measured from the center of the Brillouin zone, $\Psi_{\bf k}^\dagger$ and $\Psi_{\bf k}$ are electronic creation and annihilation operators, 
 \begin{align}
 \label{eq:h_e}
 h_{\rm e}({\bf k}) &=t_1 \sum_i \left[\cos({\bf k}\cdot{\bf a}_i) \sigma_x + \sin({\bf k}\cdot{\bf a}_i)  \sigma_y\right]\notag\\
 &- 2 t_2  \sum_i \sin({\bf k}\cdot{\bf b}_i)  \sigma_z + M \sigma_z,
 \end{align}
 $\sigma_{x,y,z}$ are Pauli matrices in the $\{A, B\}$ sublattice space of the honeycomb lattice, ${\bf a}_1 = a \hat{\bf x}$,  ${\bf a}_2 = a ( - \hat{\bf x} + \sqrt{3} \hat{\bf y} )/2$, and  ${\bf a}_3 = a( - \hat{\bf x} - \sqrt{3} \hat{\bf y})/2$
are the nearest-neighbor lattice vectors ($a$ being the lattice constant),  
${\bf b}_1= \sqrt{3} a \hat{\bf y}$, $ {\bf b}_2 = a ( -3 \hat{\bf x} - \sqrt{3} \hat{\bf y} )/2$ and  ${\bf b}_3 = a ( 3 \hat{\bf x} - \sqrt{3} \hat{\bf y})/2$
are the next-nearest neighbor lattice vectors, $t_1$ ($t_1\in\mathbb{R}$) is the nearest-neighbor hopping amplitude, $M$ is the ``Semenoff mass'' \cite{semenoff1984condensed} which breaks space inversion symmetry, and $i t_2$ is the next-nearest-neighbor hopping ($t_2\in\mathbb{R}$) leading to a ``Haldane mass''  \cite{Haldane1988} and broken time-reversal symmetry. 

\begin{figure}[t]
    \centering
    \includegraphics[width=\linewidth]{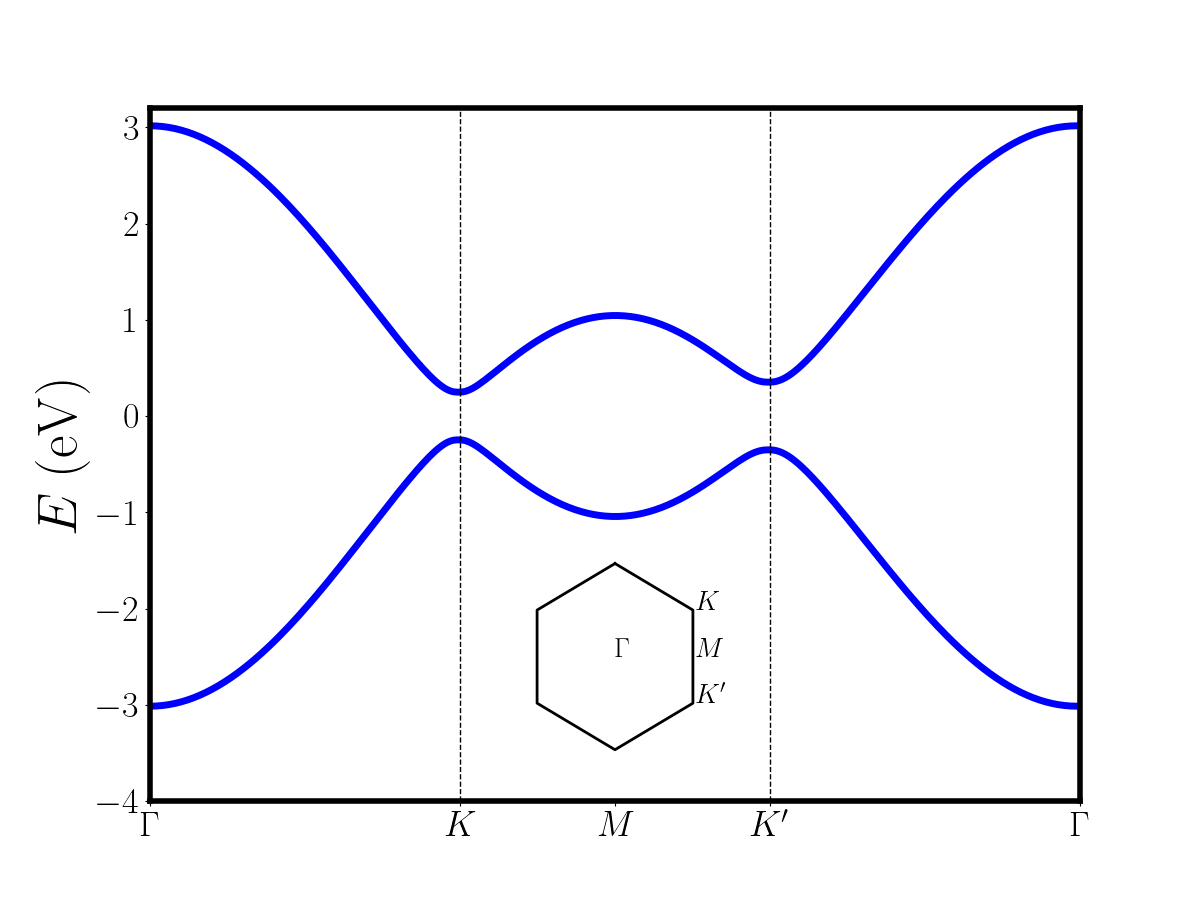}
    \caption{Electronic bands of the honeycomb lattice, with nearest-neighbor electronic hopping parameter $t_1$, Semenoff mass $M=0.3 t_1$ and second-neighbor hopping amplitude $t_2=0.01 t_1$.
   The Fermi level sits at the zero of energy. The inset shows the $K$ and $K^\prime$ valleys in the Brillouin zone.}
    \label{fig:lattice_dispersion}
\end{figure}

\begin{table}[t]
    \caption{Dirac velocities and masses of at valleys $K$ and $K'$, where $v_0=-3/2 t_1 a$. The energy gap is given by $2|m|$.}       
         \label{tab:table}
    \begin{tabular*}{\linewidth}{@{\extracolsep{\fill}} cccc }

	& $v_x$ & $v_y$ & $m$ \\
          \hline\hline
      	Valley $K$ & $v_0$ & $-v_0$ & $ M-3\sqrt{3} t_2$  \\
	\hline
       	Valley $K'$ & $v_0$ & $v_0$ & $ M+3\sqrt{3} t_2$  \\
       	 \hline
    \end{tabular*}
\end{table}

For later reference, we briefly review a few aspects of the electronic band structure. Figure \ref{fig:lattice_dispersion} shows the typical electronic bands obtained from Eq. (\ref{eq:h_e}).
Two massive Dirac fermions, each described by Eq. (\ref{eq:h0}), emerge at the two inequivalent corners (valleys $K$ and $K'$) of the hexagonal Brillouin zone. 
Their Dirac masses and velocities are  summarized in Table \ref{tab:table}.
Accordingly, the product $v_x v_y m$ has the same sign for both Dirac fermions when $|M|<3\sqrt{3}|t_2|$, but opposite sign when $|M|>3\sqrt{3}|t_2|$.
These two regimes correspond to topologically distinct electronic phases. When $|M|>3\sqrt{3}|t_2|$, the system is a Semenoff insulator and the bands have zero total Chern number (the partial Chern numbers at valleys $K$ and $K'$, each given by Eq. (\ref{eq:chern}), cancel one another upon addition).  When $|M|<3\sqrt{3}|t_2|$, the system is a Haldane/Chern insulator and the bands have nonzero total Chern number (the partial Chern numbers at valleys $K$ and $K'$ have the same value and thus do not cancel upon addition).
A topological phase transition takes place at $|M|=3\sqrt{3} |t_2|$, where the energy gap closes at one of the valleys. 
When both $M$ and $t_2$ are simultaneously nonzero, the magnitude of the energy gap (or the Dirac mass) differs at valleys $K$ and $K'$; this is known as the valley Zeeman splitting or valley Zeeman effect \cite{wang2024recent}. 
We note that the terminology ``valley-Zeeman" is also associated with a kind of spin-orbit coupling (SOC) in graphene when adsorbed on a transition metal dichalcogenide substrate \cite{morpurgo:prx, Wang2015, Ingla:2021, kumar2021, Maslov2022, mojdeh:2025}; the effect that we are discussing here is different from SOC.

Next, the electron-photon Hamiltonian for photons with negligible wave vector is given by
\begin{equation}
\label{eq:e-pt}
{\cal H}_{\rm e-pt} =\sum_{{\bf k}} \Psi_{\bf k}^\dag h_{\rm e-pt}({\bf k}) \Psi_{\bf k},
\end{equation}
where
\begin{equation}
 h_{\rm e-pt}\simeq -e {\bf A}\cdot \frac{\partial h_{\rm e}({\bf k})}{\partial {\bf k}}+\frac{e^2}{2} \sum_{\alpha,\beta}\frac{\partial^2 h_{\rm e}({\bf k})}{\partial k_\alpha \partial k_\beta} A_\alpha A_\beta
 \end{equation}
 and ${\bf A}$ is the electromagnetic vector potential.
 Equation (\ref{eq:e-pt}) simply follows from a Peierls substitution ${\bf k}\to {\bf k}-e{\bf A}$ applied to the electronic Hamiltonian.

The electron-phonon Hamiltonian requires more careful analysis. 
We first consider the coupling between electrons and in-plane lattice vibrations. 
When the phonon wave vector is neglected, only optical phonons (in which the $A$ and $B$ sublattices move out of phase) couple to electrons, through a modulation of the hopping $t_1$.
Because next-nearest neighbors connect atoms of the same sublattice, $t_2$ is not modulated by zero wave vector phonons.
Then, we get  
\begin{equation}
\label{eq:e-pn_inp}
{\cal H}_{\rm e-pn, in} =\sum_{{\bf k}} \Psi_{\bf k}^\dag h_{\rm e-pn, in}({\bf k}) \Psi_{\bf k},
\end{equation}
where (cf. Appendix A)
\begin{align}
&h_{\rm e-pn, in}({\bf k}) =\frac{1}{a}\frac{\partial t_1}{\partial a}\left({\bf u}_A-{\bf u}_B\right)\notag\\
&\cdot\sum_i {\bf a}_i\left[\cos({\bf k}\cdot{\bf a}_i)\sigma_x+\sin({\bf k}\cdot{\bf a}_i)\sigma_y\right],
\label{eq:h_e-pn}
\end{align}
and ${\bf u}_A$ and ${\bf u}_B$ are the displacements of the $A$ and $B$ sublattices with respect to their positions in the static honeycomb lattice.
It is to be noted that there is no $\sigma_z$ matrix participating in $h_{\rm e-pn, in}({\bf k})$.

Second, we consider the coupling of out-of-plane lattice vibrations to electrons. These phonons do not change the interatomic distance to first order in the atomic displacement, and therefore do not modulate $t_1$ and $t_2$ at that order (we disregard the effect of terms that are nonlinear in the lattice displacement).
Yet, out-of-plane optical phonons, in which $A$ and $B$ sublattices move out of phase, can modulate the potential difference between those sites.
As a result, the electron-phonon coupling for these phonons (at negligible wave vector) may be written as
\begin{equation}
\label{eq:e-pn_inp_2}
{\cal H}_{\rm e-pn, out} =\sum_{{\bf k}} \Psi_{\bf k}^\dag h_{\rm e-pn, out}({\bf k}) \Psi_{\bf k},
\end{equation}
where 
\begin{equation}
h_{\rm e-pn, out}({\bf k}) =g_z({\bf k}) \sigma_z ({\bf u}_A-{\bf u}_B)\cdot\hat{\bf z}
\end{equation}
and $g_z$ is a phenomenological real parameter that  can in general be ${\bf k}$-dependent.
A microscopic theory of $g_z ({\bf k})$, dependent on how out-of-plane phonons modify the potential difference between $A$ and $B$ sublattices, is beyond the scope of the present work. 
Here we instead concentrate on the constant, ${\bf k}$-independent, part of $g_z$ that amounts to a dynamical phonon-induced modulation of the Semenoff mass $M$. 
This constant is allowed to be nonzero if the crystal lacks spatial inversion symmetry (i.e. one must have $M\neq 0$).
Should inversion symmetry ${\cal P}$ be present in the crystal, the electron-phonon coupling would obey 
\begin{equation}
{\cal P} h_{\rm e-pn, out}({\bf k}) {\cal P}^{-1} = h_{\rm e-pn, out}(-{\bf k}).
\end{equation}
This condition, together with the relation ${\cal P} [({\bf u}_A-{\bf u}_B)\cdot\hat{\bf z}] {\cal P}^{-1} =  ({\bf u}_A-{\bf u}_B)\cdot\hat{\bf z}$ (${\cal P}$ interchanges $A$ and $B$ sublattices while reversing the displacement vectors) and ${\cal P}\sigma_z {\cal P}^{-1}=-\sigma_z$ (${\cal P}$ acts as $\sigma_x$ on the electronic pseudospin due to its interchange of $A$ and $B$ sites), would imply $g_z({\bf k}) = - g_z(-{\bf k})$, thereby preventing a ${\bf k}$-independent $g_z$. 

Note that there are no $\sigma_x$ and $\sigma_y$ matrices participating in $h_{\rm e-pn, out}({\bf k})$. In some sense, the in-plane and out-of-plane phonons are complementary in the way they couple to electrons. The latter modulate the electronic energy gap, while the former do not.

Finally, the change in the electron-phonon Hamiltonian due to an electromagnetic vector potential can be approximated (in the long-wavelength limit) as
\begin{equation}
{\cal H}_{\rm e-pnpt}=\sum_{{\bf k}} \Psi_{\bf k}^\dag h_{\rm e-pnpt}({\bf k}) \Psi_{\bf k},
\end{equation}
where $h_{\rm e-pnpt}({\bf k})= h_{\rm e-pn}({\bf k}-e{\bf A}) - h_{\rm e-pn}({\bf k})$.
For the calculation of the Raman tensor, it suffices to expand $h_{\rm e-pnpt}({\bf k})$ to second order in ${\bf A}$.
Then,
\begin{align}
h_{\rm e-pnpt}({\bf k}) &\simeq  -e {\bf A}\cdot \frac{\partial h_{\rm e-pn, in}({\bf k})}{\partial {\bf k}}\notag\\
&+\frac{e^2}{2} \sum_{\alpha,\beta}\frac{\partial^2 h_{\rm e-pn, in}({\bf k})}{\partial k_\alpha \partial k_\beta} A_\alpha A_\beta
\end{align}
for in-plane phonons, and
 \begin{equation}
h_{\rm e-pnpt}({\bf k}) \simeq  0
\end{equation}
for out-of-plane phonons (as we do not consider the ${\bf k}$-dependence of $g_z$).

\subsection{Raman tensor for out-of-plane phonons}

The honeycomb lattice with Haldane and Semenoff mass terms belongs to the $C_{3h}$ point group. It hosts three Raman active optical phonon modes: two doubly degenerate in-plane modes ($E_1$, $E_2$), and an out-of-plane nondegenerate mode ($A_{1}'$). In this section, we consider the Raman tensor associated to the latter.

To calculate the Raman tensor elements from microscopic theory, we take the Hamiltonians of the preceding subsection as inputs for the methodology presented in Ref.~\cite{Basko:2009} for pristine graphene. Specifically, we use Eqs. (8) and (9) from Ref. \cite{Basko:2009} and we refer the reader to our Appendix \ref{sec:efficiency} to relate the tensor ${\cal M}^{i j l}$ of Ref. \cite{Basko:2009} to our Raman tensor $R_{ij}$.  Unlike in Ref.~\cite{Basko:2009}, our theory includes $M$ and $t_2$, which play a crucial role in our results. In addition, out-of-plane phonons were not considered in Ref.~\cite{Basko:2009}.
Next, we show the outcome of the calculation.

In the backscattering configuration, the Raman tensor for the $A_1'$ mode has the form
\begin{equation}
   R_{A_1'}= \begin{pmatrix}
    R_{xx}^{\rm op} & R_{xy}^{\rm op}  \\
    -R_{xy}^{\rm op}  & R_{xx}^{\rm op} 
    \end{pmatrix},
\end{equation}
as dictated by the $C_{3h}$ point group \cite{wallis1971ionic, selcuk:2024}. 
We have added the superscript ``op'' (out-of-plane) to the Raman tensor elements, to differentiate them from those of in-plane phonons that will be discussed below.
Thus, the out-of-plane phonon has an off-diagonal part of the Raman tensor that is antisymmetric. 
This antisymmetric part vanishes when $t_2=0$, but can be very significant (comparable to the diagonal Raman tensor element) when $t_2\neq 0$, as will be shown below. 

\begin{figure}[h]
\centering
\includegraphics[width=\linewidth]{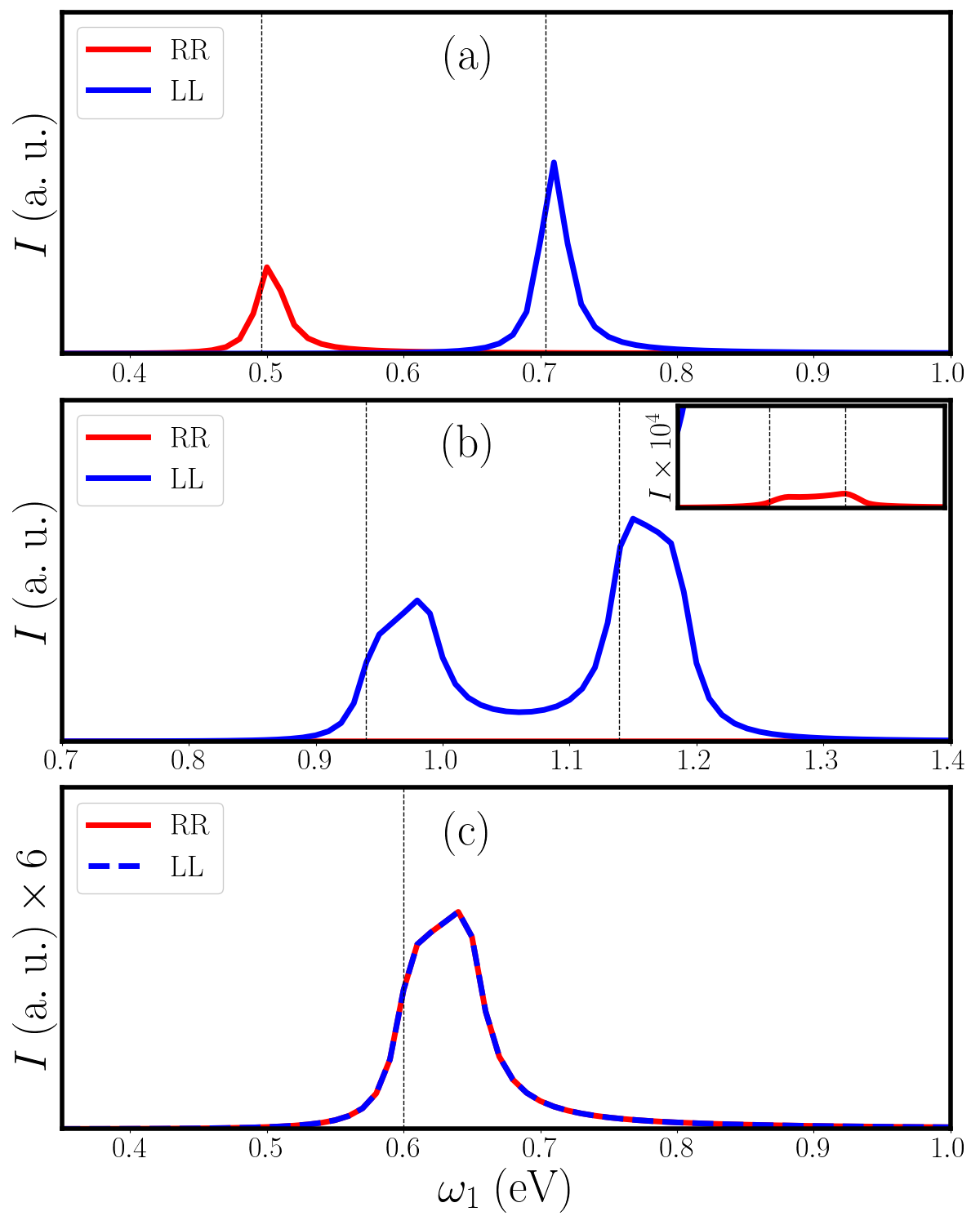}
\caption{Raman scattering intensity for an out-of-plane phonon as a function of the incident light frequency, in a tight-binding model of a honeycomb lattice with broken inversion and time-reversal symmetries. For all panels $t_1=1$ eV, $\omega_0=0.05$ eV and $\eta=0.01$ eV. The red and blue curves correspond to right- and left-circularly polarized light, respectively (same polarization for incident and detected light, denoted as $LL$ and $RR$ configuration). The vertical dashed lines are the frequencies corresponding to the two lowest electronic interband transition energies across the energy gap of the insulator, one at valley $K$ and the other at valley $K'$ a)  Semenoff insulator regime, with $t_2=0.01$ eV,  $M=0.3$ eV and a valley Zeeman splitting of $\approx 200$ meV. b) Haldane/Chern insulator regime, with $t_2=0.1$ eV,  $M=0.05$ eV and a valley Zeeman splitting of $\approx 200$ meV. Inset shows the zoom of the figure to signify a nonzero but small signal in the $RR$ configuration.  c) Semenoff insulator with time-reversal symmetry ($t_2=0$).}
\label{fig:lattice_selection}
\end{figure}

When the incident and scattered light are linearly polarized, peaks in the Raman intensity are found when $\omega_1=2 |M\pm 3\sqrt{3}t_2|$, i.e., when the incident light is in resonance with the electronic energy gaps at $K$ or $K'$ (N.B.: this statement neglects the phonon frequency. In practice, the frequency at which maximum in the intensity takes place is shifted by $\sim \omega_0$ from $2 |M\pm 3\sqrt{3}t_2|$; this shift is indeed visible in our numerical results of Fig. \ref{fig:lattice_selection}).
This is an example of the resonant Raman effect \cite{cardona2005light}. The intensity peaks are broadened by an electronic decay rate $\eta$ that we have included phenomenologically in the microscopic theory (strictly speaking, $\eta$ in our theory should be an infinitesimal positive number, as it is in Ref.~\cite{Basko:2009}. Extrapolating $\eta$ to finite values is a rough but common way to incorporate the effects of finite electron lifetimes in the Raman tensor).
The peaks corresponding to higher resonance frequencies are larger, as expected, since the Raman intensity is well-known to grow at higher frequency \cite{Cardona1982}. 

When the incident and scattered lights are circularly polarized, the results are more peculiar. Figure \ref{fig:lattice_selection} contains plots of the Raman intensity ($|\hat{\bf e}_1^\dagger\cdot {\bf R}_\lambda\cdot\hat{\bf e}_2|^2$)
when $\hat{\bf e}_1 = \hat{\bf e}_2 = (\hat{\bf x} \pm i\hat{\bf y})/\sqrt{2}$, i. e. when the incident and detected lights are circularly polarized with equal handedness (we will henceforth denote them as $RR$ and $LL$ configurations, respectively). 
In Fig. \ref{fig:lattice_selection}(a), corresponding to the Semenoff insulator regime,
an extinction of the Raman intensity is evident at the higher (lower) resonance frequency. 
This finding is consistent with the Raman selection rule unveiled in Ref. \cite{selcuk:2024} and reproduced in Eq. (\ref{eq:sel}) of the present work. 
Indeed, the extinction takes place when $\omega_1$ becomes resonant with a Dirac fermion for which $v_x v_y m$ has the appropriate sign. 
Only one extinction takes place because $v_x v_y m$ has opposite signs for the two Dirac fermions in a Semenoff insulator. 
This also proves that $R^{\rm op}_{xy}$ is as large (in modulus) as $R^{\rm op}_{xx}$ at the extinction, since 
 \begin{equation}
\label{eq:sel_latt}
|\hat{\bf e}_1^\dagger\cdot {\bf R}_\lambda\cdot\hat{\bf e}_2|^2 \propto|R^{\rm op} _{xx}|^2 |1\pm i R^{\rm op} _{xy}/ R^{\rm op} _{xx} |^2,
\end{equation}
with $\pm$ signs corresponding to $RR$ and $LL$ light configurations.

Figure \ref{fig:lattice_selection}(b) displays a similar plot, but for the Haldane/Chern insulator regime.
In this case, two extinctions are realized under the circular polarization of a given handedness, and none under circular polarization of the opposite handedness. 
This is once again consistent with Eq. (\ref{eq:sel}), as the two Dirac fermions have the same sign of $v_x v_y m$ in the Haldane insulating regime.
Finally, Fig. \ref{fig:lattice_selection}(c) displays the same quantity when $t_2=0$. Here, time-reversal symmetry is preserved, the energy gaps at $K$ and $K'$ coincide and there is no distinction between the $LL$ and $RR$ Raman spectra.
In sum, the lattice model results support Eq. (\ref{eq:ratio}), at least near resonance. 

\begin{figure}[t]
\centering
\includegraphics[width=\linewidth]{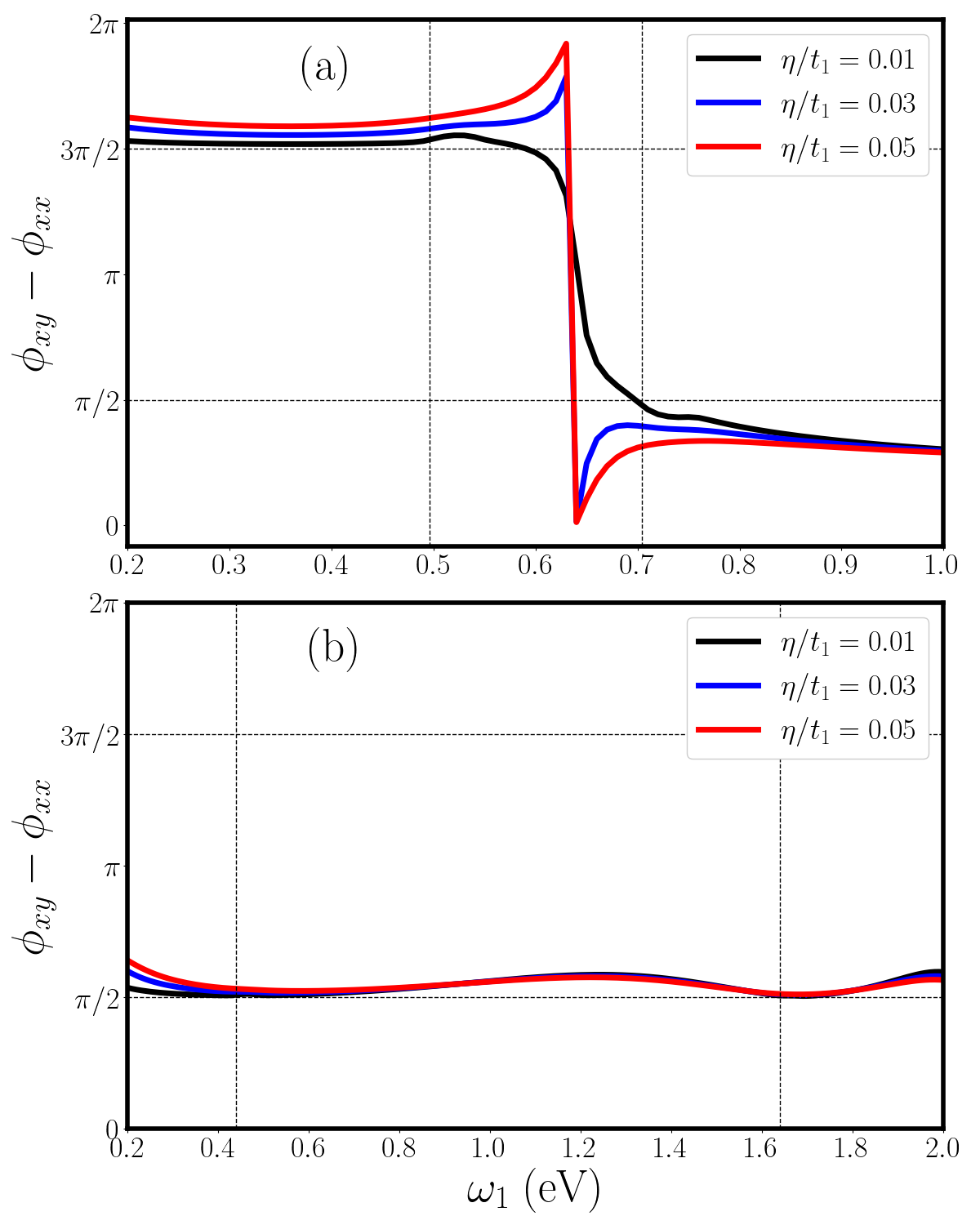}
\caption{Phase difference between Raman tensor elements $R_{xx}^{\rm op}$ and $R_{xy}^{\rm op}$ for the out-of-plane phonon in a two-dimensional honeycomb lattice with a broken inversion and time-reversal symmetry, modelled with a tight-binding electronic structure.
Different curves represent different electronic decay rate values. 
The vertical dashed lines are the frequencies corresponding to the two lowest electronic interband transition energies across the energy gap of the insulator, one at valley $K$ and the other at valley $K'$. Horizontal lines are guides to the eye.  
The phonon frequency is taken as $\omega_0=0.05$ eV. 
(a) $t_2=0.01$ eV and $M=0.3$ eV, corresponding to the Semenoff insulator regime. 
(b) $t_2=0.1$ eV and $M=0.3$ eV, corresponding to the Haldane/Chern insulator regime. }
\label{fig:lattice_phasedif}
\end{figure}
\par
Figure \ref{fig:lattice_phasedif} illustrates the phase difference $\phi_{xy}-\phi_{xx}$ as a function of $\omega_1$.
In the Semenoff insulator regime [Fig. \ref{fig:lattice_phasedif}(a)], $\phi_{xy}-\phi_{xx}$ is quantized to $-\pi/2$ (mod $2\pi$) when the frequency of the incident light is equal or less than the smallest energy gap, provided that the electronic decay rate $\eta$ is small.
Then, as $\omega_1$ varies from the first resonance to the second (higher) one, $\phi_{xy}-\phi_{xx}$ jumps by $\pi$, provided once again that $\eta$ is small. 
At even higher frequencies, quantization is lost and the phase difference varies continuously.
This result confirms that Eq. (\ref{eq:phasedif}) applies to the honeycomb lattice model, though only in the vicinity of resonances. 
For further validation, Fig.~\ref{fig:lattice_phasedif}(b)  replots the same quantity but in the Haldane/Chern insulating regime. 
In this case, the two Dirac fermions have the same sign of $v_x v_y m$. Correspondingly, analytical theory would predict that $\phi_{xy}-\phi_{xx}$ should take the same value at both resonances. This is exactly what is observed in the numerical result.
 
\subsection{Raman tensor for in-plane phonons}

Thus far we have concentrated on out-of-plane lattice vibrations, which were the subject of study in Ref. \cite{selcuk:2024}. 
It is however natural to wonder whether the Raman tensor for in-plane phonons could also harbor distinctive features in systems with 2D massive Dirac fermions. 
Elucidating this issue is the objective of the present section.

 An explicit microscopic theory calculation, following an identical approach as for out-of-plane phonons (with the caveat that some Feynman diagrams that vanished for the Raman tensor of out-of-plane phonons are no longer zero and must be taken into account), gives the following form of the Raman tensor for the in-plane (degenerate) phonons in the backscattering configuration:
\begin{align}
\label{eq:R_E}
   R_{E_2(x)} &= \begin{pmatrix}
    R^{\rm ip}_{xx} & R^{\rm ip}_{xy} \\
    R^{\rm ip}_{xy} & -R^{\rm ip}_{xx}
    \end{pmatrix}\notag\\
       R_{E_2(y)} &= \begin{pmatrix}
    R^{\rm ip}_{xy}  & -R^{\rm ip}_{xx} \\
    -R^{\rm ip}_{xx} & -R^{\rm ip}_{xy}
    \end{pmatrix},
\end{align}
as expected from the $C_{3h}$ symmetry of the crystal \cite{wallis1971ionic}.  We have added the superscript ``ip'' (in-plane) to the Raman tensor elements, to differentiate them from those corresponding to out-of-plane phonons.
We limit ourselves to $E_2$ phonons, as $E_1$ phonons are not Raman active in the backscattering configuration \cite{wallis1971ionic}.

It follows immediately from Eq. (\ref{eq:R_E}) that the Raman intensity, proportional to $|\hat{\bf e}_1^\dagger\cdot {\bf R}_{E_2(x)}\cdot\hat{\bf e}_2|^2+|\hat{\bf e}_1^\dagger\cdot {\bf R}_{E_2(y)}\cdot\hat{\bf e}_2|^2$, vanishes when $\hat{\bf e}_1 = \hat{\bf e}_2 = (\hat{\bf x} \pm i\hat{\bf y})/\sqrt{2}$.
There is therefore no selection rule between the $LL$ and $RR$ configurations. 
If we instead adopt the $LR$ or $RL$ configurations, i.e., $\hat{\bf e}_1 = (\hat{\bf x} \pm i\hat{\bf y})/\sqrt{2}$ and ${\bf e}_2={\bf e}_1^\dagger$, then we arrive at 
\begin{equation}
\label{eq:sel_latt_ip}
\sum_{\lambda\in E_2}|\hat{\bf e}_1^\dagger\cdot {\bf R}_\lambda\cdot\hat{\bf e}_2|^2 \propto|R^{\rm ip} _{xx}|^2 |1\pm i R^{\rm ip} _{xy}/ R^{\rm ip} _{xx} |^2,
\end{equation}
with $\pm$ signs corresponding to $LR$ and $RL$ light configurations. This expression is formally identical to that of out-of-plane phonons in the $LL$ and $RR$ configuration [cf. Eq. (\ref{eq:sel_latt})]. It is thus sensible to investigate whether there could be a selection rule between $LR$ and $RL$ configurations for in-plane phonons. 

\begin{figure}[h]
\centering
\includegraphics[width=\linewidth]{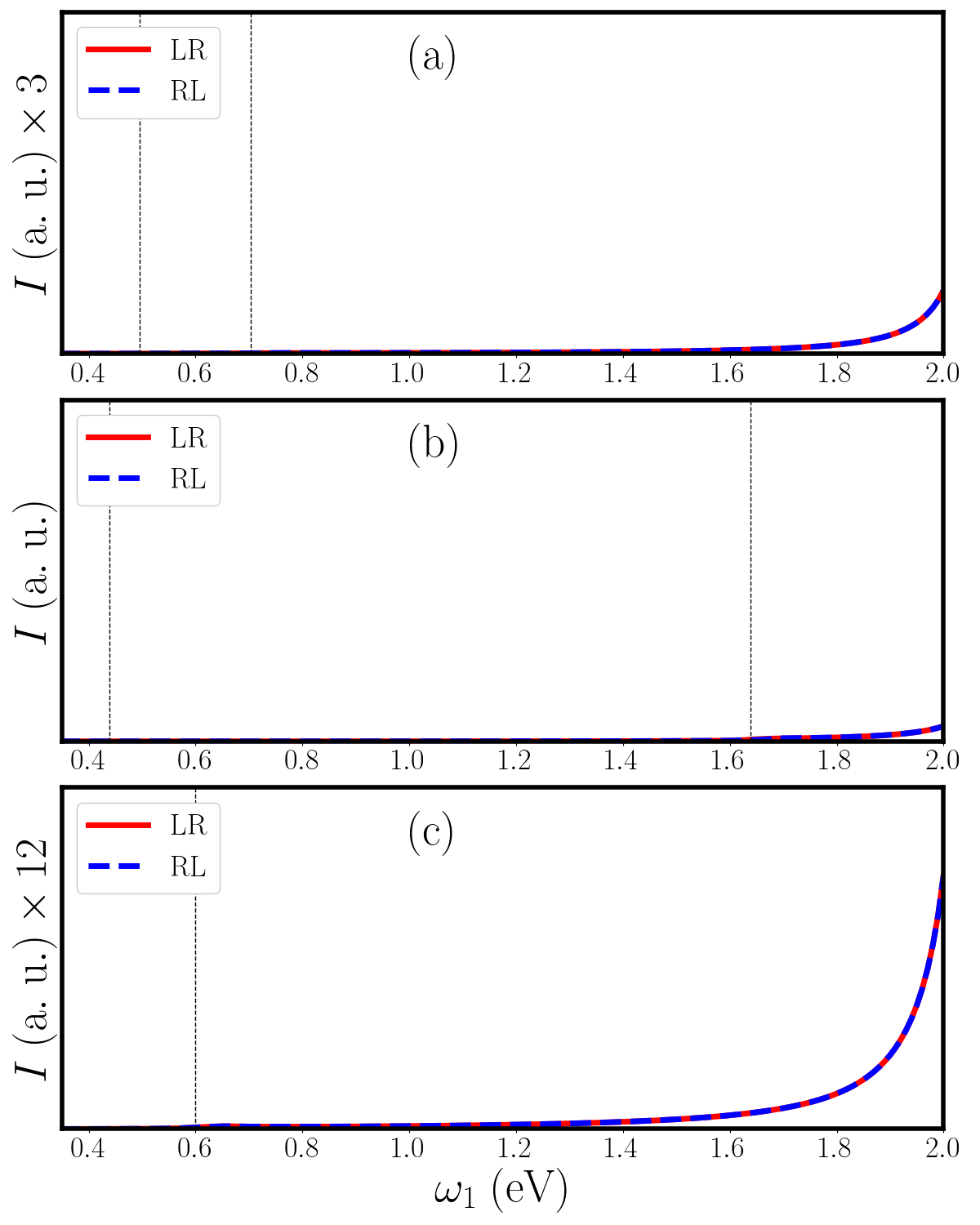}
\caption{Raman scattering intensity for the $E_2$ in-plane phonons, as a function of the incident light frequency, in a tight-binding model of a honeycomb lattice with broken inversion and time-reversal symmetries. For all panels $t_1=1$ eV, $\omega_0=0.05$ eV, $\eta=0.01$ eV and $M=0.3$ eV. The red and blue curves correspond to right- and left-circularly polarized light, respectively (opposite polarizations for incident and detected light, denoted as $LR$ and $RL$ configurations).  The vertical dashed lines are the frequencies corresponding to the two lowest electronic interband transition energies across the energy gap of the insulator, one at valley $K$ and the other at valley $K'$. a)  Semenoff insulator regime, with $t_2=0.01$ eV and a valley Zeeman splitting of $\approx 200$ meV. b) Haldane/Chern insulator regime, with $t_2=0.1$ eV and a valley Zeeman splitting of $\approx 1.20$ eV. c) Semenoff insulator with time-reversal symmetry ($t_2=0$).}
\label{fig:lattice_selection_inplane}
\end{figure}

Figure \ref{fig:lattice_selection_inplane} displays Eq. (\ref{eq:sel_latt_ip}) as a function of the incident light frequency, in the $LR$ and $RL$ configurations, for the Semenoff and Haldane insulating regimes. 
The outcome clearly shows that there are no selection rules or extinctions akin to those found for the out-of-plane phonon in Eq. (\ref{eq:sel_latt}).
The underlying reason is that $|R^{\rm ip} _{xy}|\ll |R^{\rm ip} _{xx}|$ for all values of $\omega_1$. As a result, no extinction can take place in Eq. (\ref{eq:sel_latt_ip}). For the same reason, there is very little difference in the Raman spectra for $LR$ and $RL$ configurations. 

\begin{figure}[t]
  \centering
    \includegraphics[width=\linewidth]{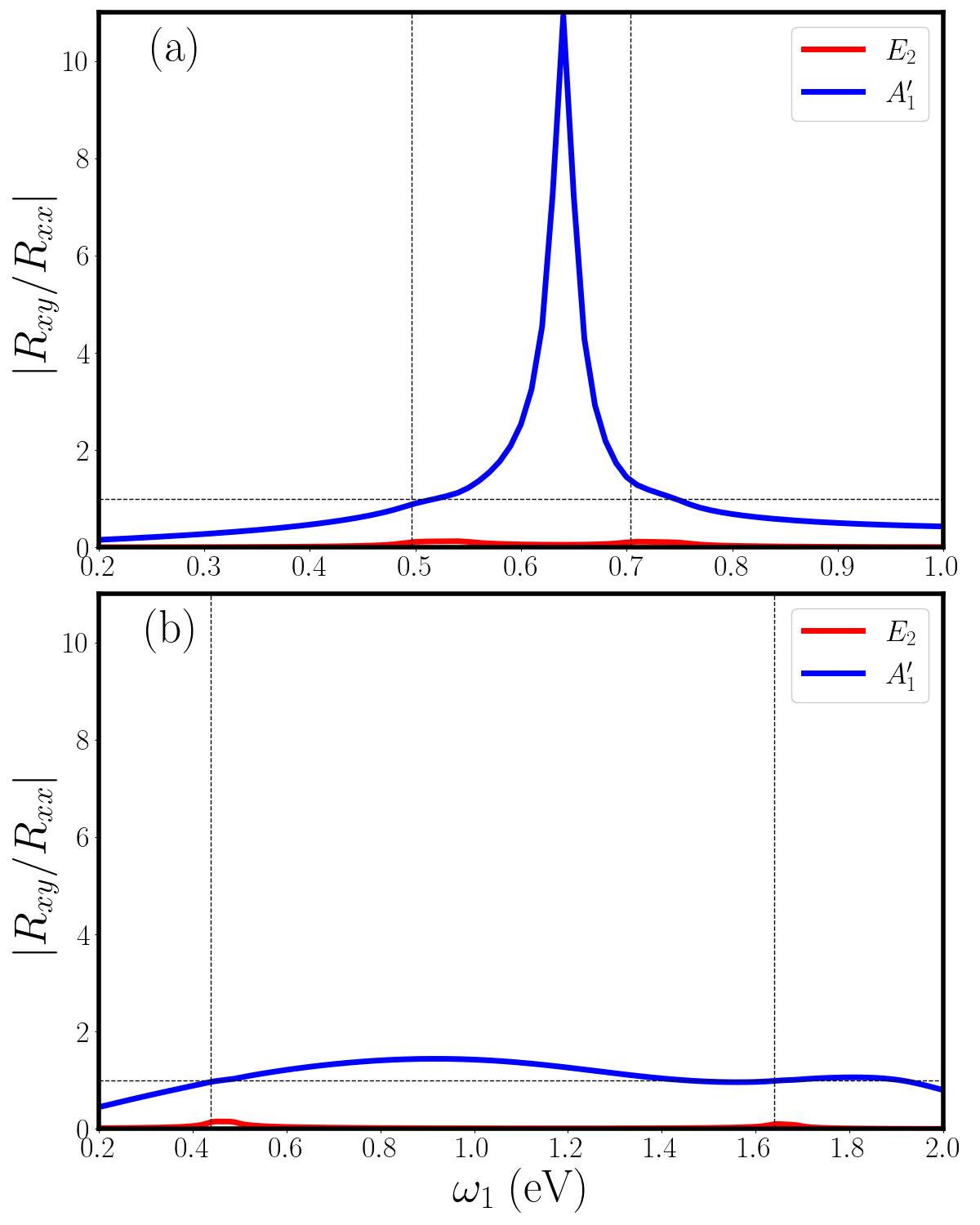}
    \caption{Ratio of the Raman tensor elements $R_{xx}$ and $R_{xy}$ for the in-plane phonon mode $E_2$ (red), and the out-of-plane phonon $A_1'$ (blue) in a two-dimensional honeycomb lattice with a broken inversion and time-reversal symmetry, modelled with a tight-binding electronic structure.
The vertical dashed lines are the frequencies corresponding to the two lowest interband transition energies across the energy gap of the insulator, one at valley $K$ and the other at valley $K'$. Horizontal dashed lines are guides to the eye for $|R_{xx}/R_{xy}|=1$.  
In both panels, $\omega_0=0.05$ eV, $t_1=1$ eV, and $\eta=0.01$ eV.
(a) $t_2=0.01$ eV and $M=0.3$ eV, corresponding to the Semenoff insulator regime. 
(b) $t_2=0.1$ eV and $M=0.3$ eV, corresponding to the Haldane/Chern insulator regime.  
    }
   \label{fig:rxyrxxratio}
\end{figure}

The reason why $|R^{\rm ip} _{xy}|\ll |R^{\rm ip} _{xx}|$ is interesting. It originates from the fact that the electronic model is invariant under the ${\cal S}={\cal M}_y {\cal T}$ symmetry, where ${\cal M}_y$ is a mirror plane perpendicular to the $y$ axis and ${\cal T}$ is the time-reversal operator. 
As shown in Appendix \ref{sec:sym}, when the phonon frequency is neglected, the ${\cal S}$ symmetry imposes $R^{\rm ip}_{xy}=-R^{\rm ip}_{yx}$ for the $E_2(x)$ phonon, and $R^{\rm ip} _{xy}=R^{\rm ip} _{yx}$ for $E_2(y)$. Comparing with the form imposed by $C_{3h}$ symmetry in Eq.~\ref{eq:R_E}, we have $R_{xy}^{\rm ip}=0$ when the phonon frequency is neglected. There is no such constraint on  $R^{\rm ip} _{xx}$, or even on $R^{\rm op} _{xy}$. Since realistic phonon frequencies constitute a small energy scale in the problem, whose neglect in the theory usually brings about relatively minor quantitative changes, $|R^{\rm ip} _{xy}|$ remains small even when $\omega_0$ is not neglected. This state of affairs is illustrated in Fig. \ref{fig:rxyrxxratio}. This figure also shows that, for out-of-plane phonons, $|R^{\rm op} _{xy}|\simeq |R^{\rm op} _{xx}|$ at the resonances, in agreement with Eq. (\ref{eq:ratio}) and thereby enabling the extinctions in Eq. (\ref{eq:sel_latt}). 

Incidentally, the absence of marked intensity peaks at the resonance frequencies in Fig. \ref{fig:lattice_selection_inplane}  is curious. 
We attribute this to the fact that $R^{\rm ip}_{xx}$ vanishes when we use the continuum approximation of Eq. (\ref{eq:h0}), as a consequence of rotational symmetry. Indeed, higher order (warping) terms in wave vector need to be added to the model of  Eq. (\ref{eq:h0}) to get a nonzero $R^{\rm ip}_{xx}$ from Dirac fermions, as first pointed out in Ref. \cite{Basko:2009}. No such requirement applies to out-of-plane phonons. Thus, the main contribution to $R^{\rm ip}_{xx}$ comes from higher energy states (away from the band-edge at $K$ and $K'$ valleys). In other words, the expected intensity peaks when the light frequency matches the energy gaps are suppressed.

\begin{figure}[t]
  \centering
    \includegraphics[width=\linewidth]{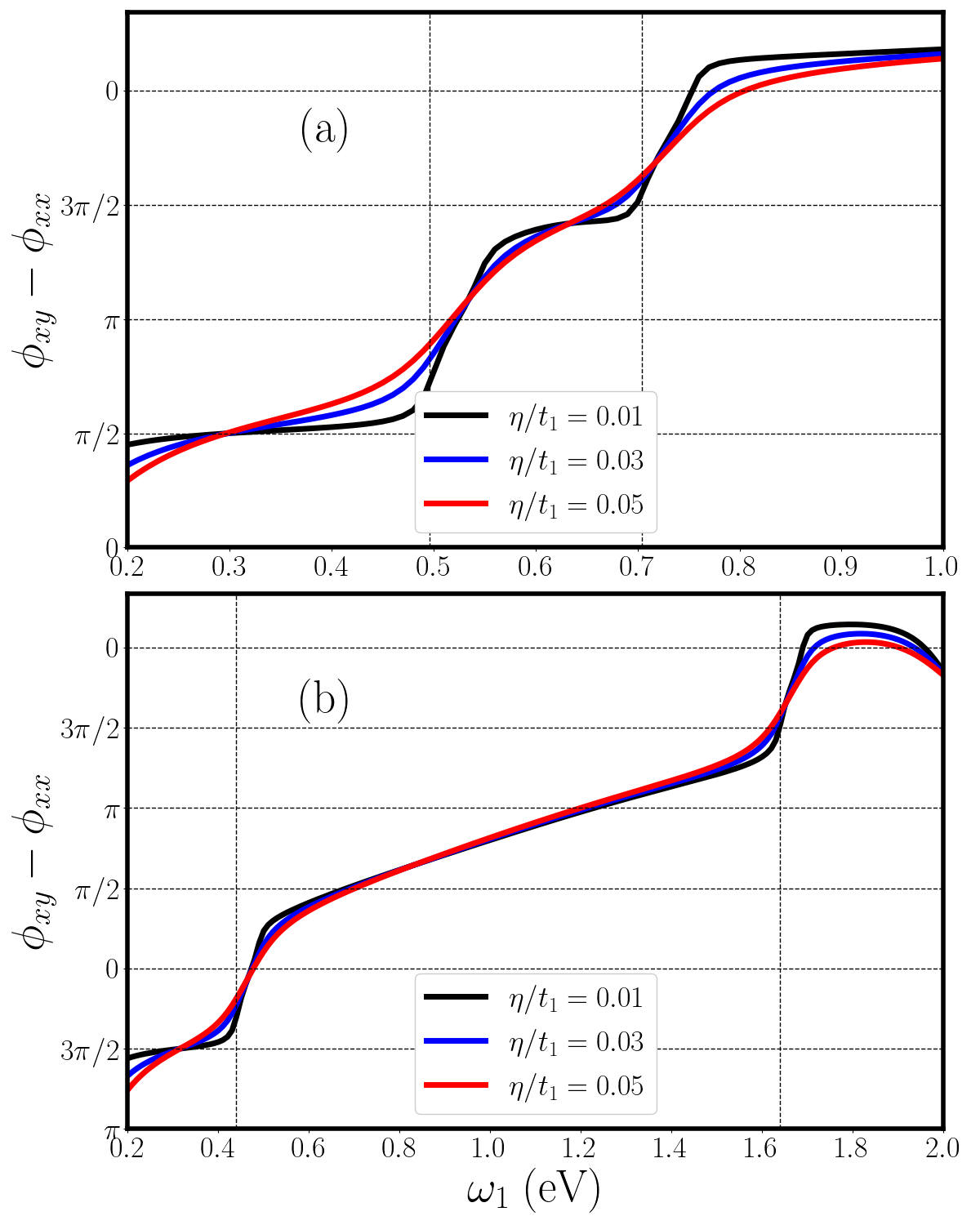}
    \caption{Phase difference between Raman tensor elements $R_{xx}^{\rm ip}$ and $R_{xy}^{\rm ip}$ for the in-plane phonon mode $E_2$ in a two-dimensional honeycomb lattice with a broken inversion and time-reversal symmetry, modelled with a tight-binding electronic structure.
Different curves represent different electronic decay rate values. 
The vertical dashed lines are the frequencies corresponding to the two lowest interband transition energies across the energy gap of the insulator, one at valley $K$ and the other at valley $K'$. Horizontal dashed  lines are guides to the eye.  
The phonon frequency is taken as $\omega_0=0.05$ eV. 
(a) $t_2=0.01$ eV and $M=0.3$ eV, corresponding to the Semenoff insulator regime. 
(b) $t_2=0.1$ eV and $M=0.3$ eV, corresponding to the Haldane/Chern insulator regime. 
    }
   \label{fig:phasedif_inplane}
\end{figure}
\par 
Now that a selection rule for in-plane phonons has been ruled out, we may ask what happens to the phase difference. The $E_2$ phonons have two independent Raman tensor elements, $R_{xx}^{\rm ip}$ and $R_{xy}^{\rm ip}$. One can therefore define (and in principle measure) a phase difference $\phi_{xy}-\phi_{xx}$.
Figure~\ref{fig:phasedif_inplane} illustrates $\phi_{xy}-\phi_{xx}$ as a function of $\omega_1$.
The ressemblance with the out-of-plane phonon case stands out.
In the Semenoff insulator regime [Fig.~\ref{fig:phasedif_inplane}(a)], $\phi_{xy}-\phi_{xx}$ changes by $\pi$ as $\omega_1$ varies from the first resonance to the second (higher) one,  for small $\eta$ values.  In the Haldane/Chern insulating regime  [Fig.~\ref{fig:phasedif_inplane}(b)], $\phi_{xy}-\phi_{xx}$ takes the same value at both resonances. 
In sum, $\phi_{xy}-\phi_{xx}$ displays a similar trend for in-plane and out-of-plane phonons, despite the fact that Eq. (\ref{eq:ratio}) does not at all apply to in-plane phonons (as evidenced by Fig. \ref{fig:rxyrxxratio}).

The preceding result can be validated from a continuum model approximation near the $K$ and $K'$ valleys.
Let $R^{\rm ip}_{ij; v}$ be the contribution to the Raman tensor element $(i, j)$ from the electronic states near the valley $v$.
It is reasonable to approximate $R^{\rm ip}_{ij}\simeq R^{\rm ip}_{ij; v}$ when the incident light frequency matches the energy gap at valley $v$.
A calculation based on the continuum approximation gives
\begin{equation}
\label{eq:ratio_ip}
R^{\rm ip}_{xy; v}/R^{\rm ip}_{xx; v}\propto \omega_0 {\rm sgn}(m_v) (-1)^v,
\end{equation}
where $m_v$ is the Dirac mass at valley $v$ (i.e., $m_K=M-3\sqrt{3} t_2$ and $m_{K'}=M+3\sqrt{3} t_2$), and $(-1)^v$ takes opposite signs for valleys $K$ and $K'$.
The proportionality factor in Eq. (\ref{eq:ratio_ip}) is not simple to obtain analytically, largely because hexagonal warping must be included in the analysis (otherwise $R^{\rm ip}_{ij; v}$ would vanish). We have nevertheless checked the veracity of Eq. (\ref{eq:ratio_ip}) numerically.
In the Semenoff insulating regime, where ${\rm sgn}(m_K) ={\rm sgn}(m_{K'})$,  Eq. (\ref{eq:ratio_ip}) has opposite signs in the two valleys. 
Therefore, the value of $\phi_{xy}-\phi_{xx}$ differs by $\pi$ at the two resonant frequencies. 
In the Haldane insulating regime, where ${\rm sgn}(m_K) =-{\rm sgn}(m_{K'})$, Eq. (\ref{eq:ratio_ip}) has the same sign in the two valleys. 
Therefore, $\phi_{xy}-\phi_{xx}$ takes the same value at the two resonant frequencies. 

Finally, the fact that $|R_{xy}^{\rm ip}|$ is very small makes it more difficult to experimentally observe $\phi_{xy}-\phi_{xx}$, since the term in the Raman intensity that depends on $\phi_{xy}-\phi_{xx}$  is proportional to the product $|R_{xx}^{\rm ip}| |R_{xy}^{\rm ip}|$ \cite{selcuk:2024}. In view of this and in view of the absence of selection rule for in-plane phonons, our focus for the remainder of this work will be on out-of-plane phonons.



\section{First-principles calculation}
\label{sec:first}
\begin{figure}[t]
  \centering
    \includegraphics[width=\linewidth]{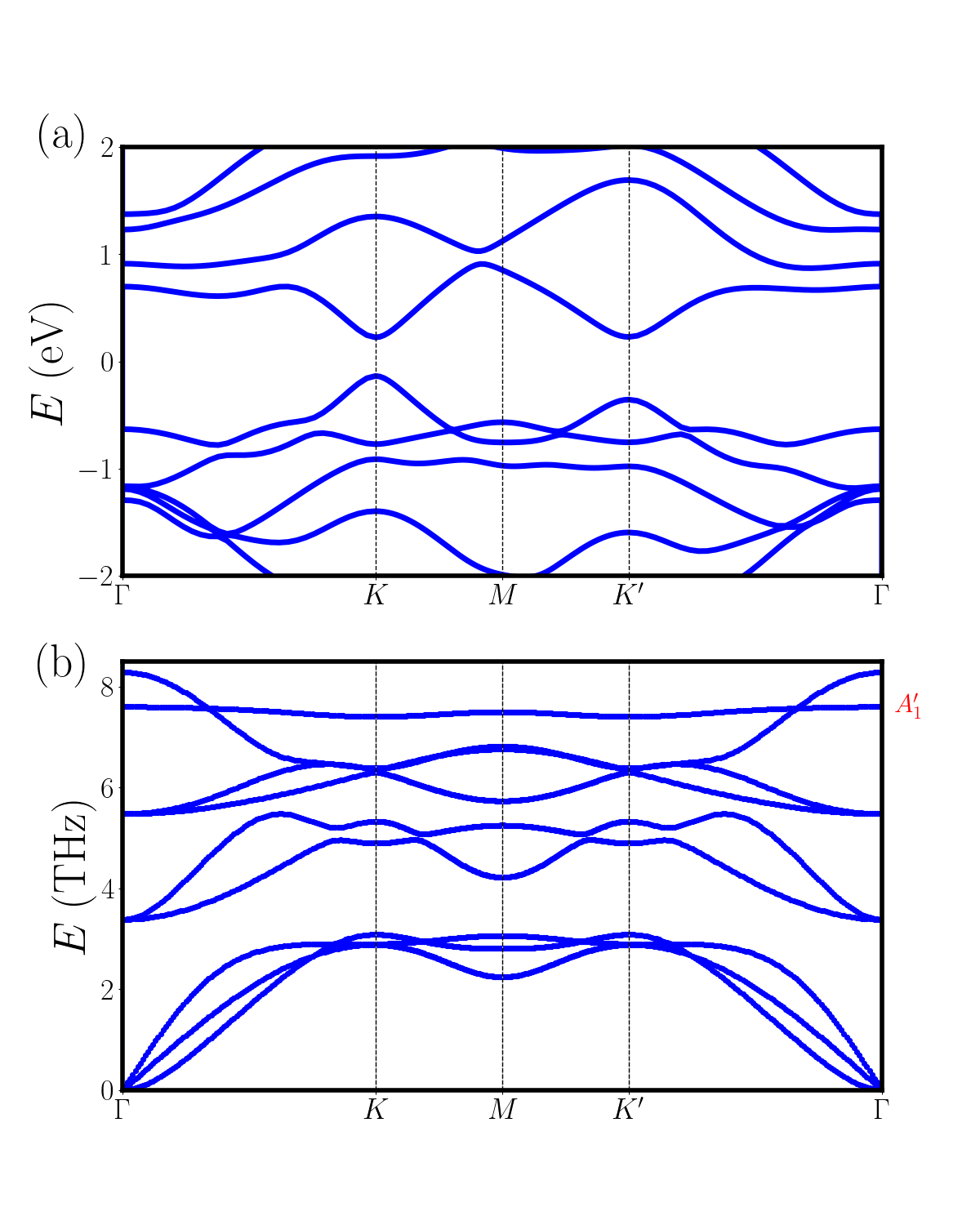}
    \caption{(a) Electronic band structure of monolayer 2H-RuCl$_2$, obtained from density functional theory. The $K$ and $K^\prime$ valleys host massive Dirac fermions with band gaps $0.36$ eV and $0.58$ eV. The valley Zeeman splitting of $\simeq 0.22 {\rm eV}$ matches well with results from Ref.~\cite{runhan:2023}. (b) Phonon band structure of monolayer 2H-RuCl$_2$, obtained from density functional theory. In the present study, we are interested in the Raman tensor for the out-of-plane optical phonon mode of symmetry $A_1'$ at the $\Gamma$ point (indicated at the upper right corner of the figure).
    }
   \label{fig:dft_band_structure}
\end{figure}

  In this section, we perform first-principles calculations of the Raman tensor on monolayer 2H-RuCl$_2$.
 This crystal with honeycomb structure has been recently predicted \cite{runhan:2023} to be a ferromagnetic insulator with out-of-plane magnetization, where low-energy electronic excitations are massive Dirac fermions located at valleys $K$ and $K^\prime$ [cf. Fig. \ref{fig:dft_band_structure}(a)].  The ferromagnetic order breaks time reversal symmetry and produces a large valley Zeeman splitting of $\approx 200$ meV. As a result, the system is an propitious testing bed for the analytical predictions of Ref. \cite{selcuk:2024}.

We have performed first-principles Raman tensor calculations using the Vienna \textit{Ab initio} Simulation Package (VASP) and the PHONOPY package \cite{vasp1,vasp2,phonopy1,phonopy2}. A DFT+U calculation has been carried out with $U=2$ eV for the Coulomb interaction, following Ref.~\cite{runhan:2023}. A $k$ mesh of $12\times12\times1$ has been selected for cell structure optimization and self-consistent calculation. The Perdew-Burke-Ernzerhof exchange correlation functional has been used throughout the calculation. A $4\times 4\times1$ supercell has been created by PHONOPY and all calculations have been done with  an energy cutoff of $500$ eV and energy tolerance of $10^{-7}$ eV. 
The crystal structure was fully relaxed until the forces on each atom were smaller than $0.001\,{\rm eV}$. 
The phonon band structure shown in Fig. \ref{fig:dft_band_structure}b  has been
obtained by a density-functional perturbation theory calculation, and phonon modes have been extracted by PHONOPY.

A finite difference scheme has been used to calculate the Raman tensor for the phonon mode $\lambda$ with negligible wave vector, through the relation \cite{refphonopy}
\begin{equation}
\label{eq:R_alter}
    R_{\alpha\beta} =\frac{\partial \epsilon_{\alpha \beta}(\omega,u_\lambda)}{\partial u_\lambda},
    \end{equation}
where $\epsilon(\omega,u_\lambda)$ is the dynamical dielectric tensor at frequency $\omega$ and $u_\lambda$ is the normal coordinate for mode for the mode $\lambda$. Displaced atomic configurations in mode $\lambda$ are extracted from PHONOPY, and dynamical dielectric functions are calculated by VASP with a $k$ mesh of $80 \times 80\times 1$ to achieve convergence. 
In Eq.~(\ref{eq:R_alter}), the frequency $\omega$ can be identified to that of the incident light. 
The effect of a finite phonon frequency is neglected. 
While this may lead to quantitative error in the vicinity of resonances, our calculations of Sec. \ref{sec:tight} indicate that finite phonon frequency effects do not greatly alter the results.

In this work, we are interested in the out-of-plane optical mode at zero wave vector. Consistent with symmetry arguments, the numerical computation of Eq. (\ref{eq:R_alter}) produces a Raman tensor of the type
\begin{equation}
   R_{A_1'}= \begin{pmatrix}
    R_{xx} & R_{xy} \\
    -R_{xy} & R_{xx}
    \end{pmatrix}
  \end{equation}
in the backscattering configuration.

\begin{figure}[t]
    \centering
\includegraphics[width=\linewidth]{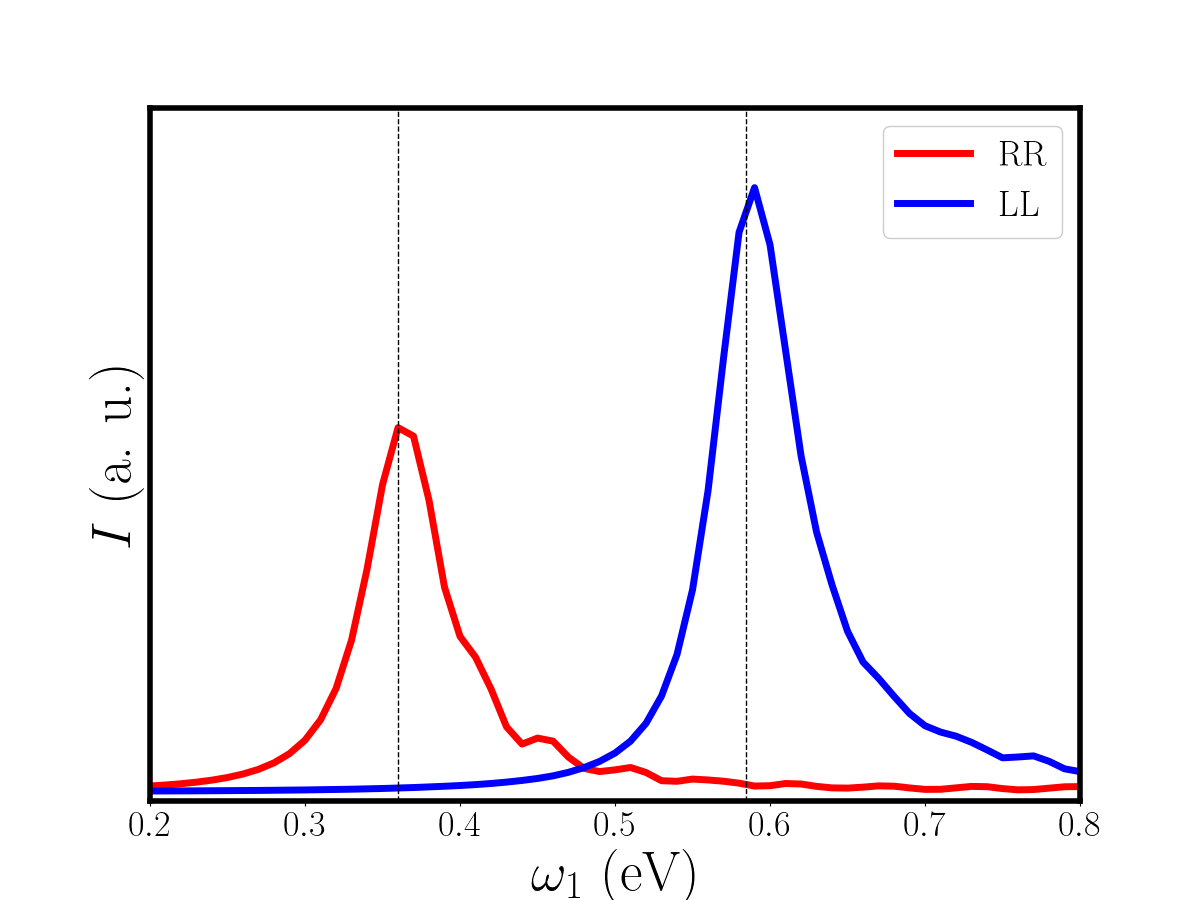}
    \caption{Raman scattering intensity for the out-of-plane $A_1'$ phonon in monolayer 2H-RuCl$_2$, calculated from first-principles, when the incident and detected lights are circularly polarized with the same handedness ($LL$ and $RR$ configurations). The dashed vertical lines represents the lowest resonant Raman frequencies in which the photon energy matches the interband transition energy across the energy gap at valleys $K$ and $K'$, respectively. The electronic decay rate is $\eta=0.03$ eV.}
    \label{fig:dft_selection}
\end{figure}

We are now ready to discuss the numerical results. We start by evaluating the Raman intensity when the incident and scattered lights are circularly polarized, with the same handedness (Fig.~\ref{fig:dft_selection}). 
An extinction of the Raman intensity is observed at one of two the resonant frequencies corresponding to the energy gaps at $K$ or $K'$; which one of the two resonances undergoes an extinction depends on whether the light configuration is $LL$ or $RR$.
This observation matches well with the results in lattice model in Fig.~\ref{fig:lattice_selection}, and also with the  continuum model results from Ref.~\cite{selcuk:2024}. Indeed, there is only one extinction per light polarization because the valley Zeeman splitting in 2H-RuCl$_2$ is not sufficiently strong to induce a band inversion; as such, the material is a Semenoff insulator with opposite signs of $v_x v_y m$ for the Dirac fermions at $K$ and $K'$.


\begin{figure}[t]
    \centering
\includegraphics[width=\linewidth]{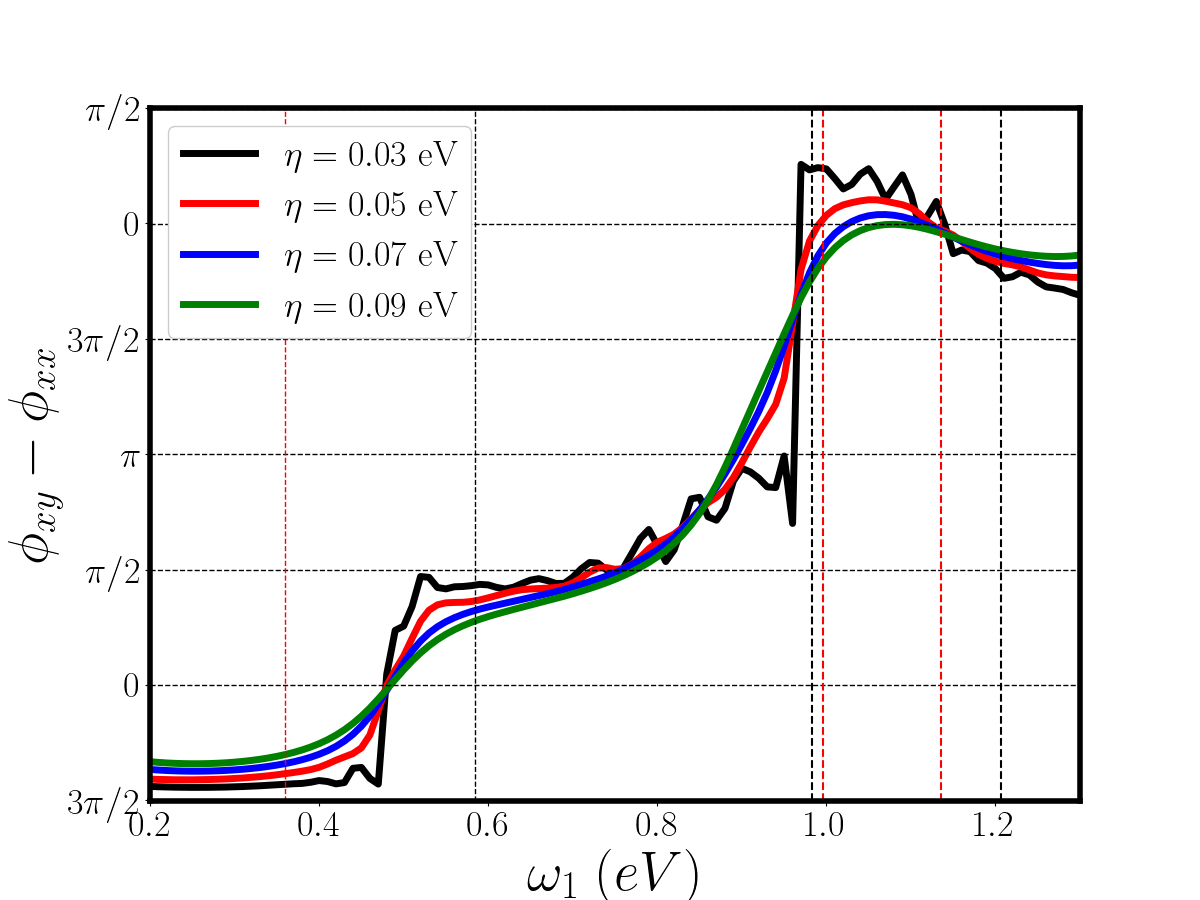}
    \caption{Phase difference between Raman tensor elements $R_{xx}$ and $R_{xy}$ for the out-of-plane $A_1'$ phonon in monolayer 2H-RuCl$_2$, calculated from first principles. Different solid lines represent various electronic decay rate values. Red and black vertical dashed lines represent interband transition energies across the bandgap in the  $K$ and $K^\prime$ valleys, respectively. Horizontal dashed lines are guides to the eye.}
    \label{fig:dft_phasedif}
\end{figure}

Next, we turn to Fig.~\ref{fig:dft_phasedif}, which shows the phase difference $\phi_{xy}-\phi_{xx}$ as a function of $\omega_1$.
The smallest value chosen for the electronic decay rate $\eta$ is  $0.03$ eV; lower values lead to difficulties to reach $k$-mesh numerical convergence. 
For $\eta=0.03$ eV, the figure clearly shows a step change in $\phi_{xy}-\phi_{xx}$ by $\pi$ as a function of $\omega_1$, from approximately $-\pi/2$ at the first resonance to approximately $\pi/2$ at the second resonance. These two resonance frequencies correspond to the lowest energy interband transitions across the bandgap at valleys $K$ and $K'$, respectively. We attribute the slight deviation of $\phi_{xy}-\phi_{xx}$ from $\pm \pi/2$ at the resonances to the fact that $\eta$ is sizable. 

When the value of $\eta$ increases, the quantization of $\phi_{xy}-\phi_{xx}$ near the aforementioned resonances is gradually washed out. This trend is also seen in the lattice model calculations of Sec.~\ref{sec:tight} and in the continuum model calculation of Ref.~\cite{selcuk:2024}. 

As $\omega_1$ exceeds the second resonant frequency, the phase difference  begins to depart from $\pi/2$. In this regime, the Raman tensor is not dominated by the contribution from a single Dirac fermion, and as a result Eq. (\ref{eq:phasedif}) is not expected to hold. 
This too follows the trend of the lattice model calculations of Sec.~\ref{sec:tight} and the continuum model calculation of Ref.~\cite{selcuk:2024}. 

At still higher frequencies, other resonances are reached when the photon frequency matches interband transition energies. 
Yet, those transitions do not connect bands described by a Dirac Hamiltonian. As a result, Eq. (\ref{eq:phasedif}) does not apply and indeed the numerical result shows variations of the phase difference that are not consistent with the analytical theory. 

There is, however, a curious aspect at the resonance frequencies near $\omega_1\simeq 1{\rm eV}$, when the interband transitions connect a Dirac-like band with a non-Dirac-like band, first in one valley and then in the other (see Fig.~\ref{fig:dft_band_structure}a). Near these two resonances, the phase difference $\phi_{xy}-\phi_{xx}$ changes abruptly by $\simeq \pi/2$. We have no analytical explanation for this behavior, aside from the speculative guess that $\pi/2$ is half of $\pi$ because only one out of the two bands connected by photons of $\simeq 1$ eV is Dirac-like. 
At even higher frequency resonances, when photons connect two non-Dirac-like bands, $\phi_{xy}-\phi_{xx}$ does not show any remarkable features.


\par 
 \section{Summary and conclusions}
\label{sec:conc}

The present study was motivated by Ref. \cite{selcuk:2024}, where two peculiar features were predicted in the Raman response for out-of-plane phonons in 2D materials hosting massive Dirac fermions. First, the phase difference between diagonal and off-diagonal Raman tensor elements was found to be quantized to $\pm \pi/2$, sensitive only to the product of the signs of the Dirac fermion masses and velocities.  Second, a selection rule was identified in the Raman intensity under circularly polarized light, leading to extinctions of the Raman intensity at certain resonant frequencies, depending once again on the product of the signs of the Dirac fermion masses and velocities.

Those predictions of  Ref. \cite{selcuk:2024} were based on a low-energy effective model in the continuum approximation, where results could be obtained analytically. In the present work, we have assessed the robustness of the analytical results in more realistic settings that reach far beyond the two-band toy model that was employed in Ref. \cite{selcuk:2024}. First, we have calculated the Raman tensor using an electronic tight-binding model on a honeycomb lattice with broken time-reversal and inversion symmetry.
Second, we have computed the Raman tensor from density functional theory for a monolayer of ferromagnetic 2H-RuCl$_2$.
Both calculations have corroborated the analytical results found in the continuum model, provided that the incident light frequency is close to the energy gap of the Dirac fermion. 

 As another novelty with respect to Ref. \cite{selcuk:2024}, we have also considered the Raman tensor for in-plane phonons.
For these, we have noted that there is no selection rule akin to the one found for out-of-plane phonons.  This feature further distinguishes our predicted Raman selection rule from the standard valley selection rule found in optical conductivity, because for one thing the latter is independent of phonons. 
In contrast, the phase difference between Raman tensor elements for in-plane phonons displays similar characteristics as those for out-of-plane phonons. At any rate, observing the phase-difference quantization for in-plane phonons will be much more difficult than for out-of-plane phonons, because in the former case one of the two key Raman tensor elements vanishes in the limit of negligible phonon frequency.

There are challenges to overcome to be able to validate or refute the preceding theoretical predictions in experiment.
First, continuously tuneable laser frequencies are required.
Second, 2D Dirac materials with large bandgaps (meaning higher Raman scattering intensities near resonance) and valley Zeeman splittings exceeding the electronic decay rate are needed.
Valley Zeeman splitting can be induced by external magnetic fields \cite{li2014valley}, by proximity to magnetic overlayers \cite{qi2015giant}, or by intrinsic magnetism in the 2D material of interest \cite{runhan:2023}. The last option is probably the more desirable in practice from the point of view of simplicity. In addition, the valley Zeeman splitting induced by external magnetic fields is weak ($\lesssim 1 {\rm meV}$/Tesla; see e.g. Ref. \cite{arora2016valley}), whereas large ($\gtrsim 100$ meV) valley Zeeman splitting from intrinsic out-of-plane ferromagnetism has been predicted by DFT in monolayer MX$_2$ materials (M=Ru, Os; X=Cl, Br) \cite{xie2024giant}.  
Unfortunately, the energy gaps of the latter materials are about a factor of two smaller than $1$ eV, while standard Raman laser frequencies exceed $1$ eV. In the short term, transition metal dichalcogenide monolayers on magnetic insulator substrates may offer more viable platforms to test our predictions. 
For instance, in monolayer WS$_2$, whose energy gap exceeds $1$ eV, a proximity coupling to an insulating magnet (EuS) has been reported to lead to a valley Zeeman splitting of $16\,{\rm meV}$ under a perpendicular magnetic field of $1$ T \cite{norden2019giant}. The magnetic field would be dispensable if the easy axis of the magnet were perpendicular to the transition metal dichalcogenide.


On the theoretical side, one could perform a first-principles calculation of the Raman tensor from the same formalism as the one employed in Sec. \ref{sec:tight} and in Refs. \cite{Basko:2009, selcuk:2024}, instead of using Eq. (\ref{eq:R_alter}).
This would enable a better consideration of finite phonon frequency effects in the DFT approach.
Additional future directions of research would be to generalize our theory to multilayer 2D Dirac materials, as well as to 3D Dirac materials or other unconventional materials like Weyl semimetals (see Ref.~\cite{drichko2025probing} for a recent example), in order for example to find out whether the peculiar band structures of these systems can leave fingerprints in the phases of the Raman tensor elements. Finding an elegant explanation for the simplicity of Eq. (\ref{eq:ratio}), if one exists,  would likewise be desirable.
\acknowledgements
S. P and I. G. have been financially supported by the Natural Sciences and Engineering Research Council of Canada (Grant No. RGPIN-2024-05210). A.K. acknowledges support from Canada First Research Excellence Fund and the Natural Sciences and Engineering Research Council of Canada (Grant No. RGPIN-2019-05312). 
M.G.V. and R.L. acknowledge the support of the Canada Excellence Research Chairs Program for Topological Quantum Matter.
The authors benefited from their affiliation to the RQMP, a strategic research network funded by the Fonds de Recherche du Québec, https://doi.org/10.69777/309032. I.G. is grateful for the hospitality of the Donostia International Physics Center while some of the present work was being conducted. I.G. and S. P. acknowledge informative discussions with C.-T. Chen, R. Martel and M. Pimenta.
\appendix
\begin{widetext}
\section{Coupling of electrons to in-plane phonons in the honeycomb lattice}
This appendix contains the derivation of Eq.~(\ref{eq:h_e-pn}). 
The nearest-neighbor hopping appearing in Eq. (\ref{eq:h_e}) is a function of the interatomic distance. Explicitly, we may write the nearest-neighbor term as
\begin{equation}
\label{eq:he_app}
 h_{\rm e}^{t_1}({\bf k}) = \sum_i t_1(|{\bf a}_i|) \left[\cos({\bf k}\cdot{\bf a}_i) \sigma_x + \sin({\bf k}\cdot{\bf a}_i)  \sigma_y\right].
\end{equation}
In the absence of phonons, $t_1(|{\bf a}_i|)=t_1(a)$ for all $i$.
In-plane optical phonons of zero wave vector modulate the nearest-neighbor hopping amplitude,
\begin{equation}
\label{eq:t1mod}
t_1(|{\bf a}_i+({\bf u}_A-{\bf u}_B)|)\simeq t_1(a)+\frac{1}{a}\frac{\partial t_1(a)}{\partial a} ({\bf u}_A-{\bf u}_B)\cdot{\bf a}_i,
\end{equation}
to first order in the atomic displacements ${\bf u}$.
Replacing the second term of Eq. (\ref{eq:t1mod}) in Eq. (\ref{eq:he_app}), we arrive at Eq.~(\ref{eq:h_e-pn}).

\section{Definition of the Raman efficiency}
\label{sec:efficiency}

This appendix explains the definition of the Raman efficiency in Eq.~(\ref{eq:rint}). Its objective is to relate the quantity we calculate to the ones that have been calculated earlier \cite{loudon1963, Basko:2009}.

Suppose an initial state that consists of a crystal in its ground state,  $n_1$ incident photons, zero outgoing photons and $n_0$ optical phonons of a mode $\lambda$. The transition probability to a final state with the crystal in the ground state, $n_1-1$ incident photons, $1$ outgoing photon, and $n_0+1$ optical phonons can be obtained from Fermi's golden rule. 
Following Ref.~\cite{loudon1963},  the transition probability after a time $\tau$ is
\begin{equation}
\label{eq:S_prob}
S_\lambda(\tau)=\tau \frac{4 \pi^3e^4 n_1(n_0+1)}{\epsilon_r^2 MN \omega_0 \omega_1 \omega_2}  \sum_{\boldsymbol{\kappa},{\bf k}_2} \left |R_{ijl}  (e_1^i)^* e_2^j e^l_{\lambda} \right |^2\delta(\omega_1-\omega_2-\omega_0)\frac{(2 \pi)^3}{V}\delta({\bf k}_1-{\bf k}_2-\boldsymbol{\kappa}),
\end{equation}
where a summed over repeated indices $i, j, l\in\{x, y, z\}$ is implicit, $e$ is the electric charge, $M$ is the reduced atomic mass for the two countervibrating sublattices, $N$ is the number of unit cells in the crystal, $\epsilon_r$ is the relative dielectric constant of the material,  ${\bf k}_1$, ${\bf k}_2$, and $\boldsymbol{\kappa}$ are the wave vectors of incident photons, scattered photons and the phonon, respectively, $V=L_x L_y L_z$ is the volume filled by the electromagnetic waves, ${\bf e}_1$, ${\bf e}_2$ and ${\bf e}_{\lambda}$ are polarizations of incident photons, scattered photons and the phonon, respectively.
Finally, $R_{ijl}$ is the transition matrix element identical to the one defined in Ref.~\cite{loudon1963}, modulo the facts that (i) we write velocity operators instead of momentum operators, (ii) we take out the polarization components $e_1^i$, $e_2^j$ and $e^l_{\lambda}$ outside the expression of $R_{ijl}$, and (iii) we include all the amplitudes described in Ref.~\cite{Basko:2009}. 


To describe the Raman scattering efficiency, first we convert and evaluate the summations in Eq. (\ref{eq:S_prob}) into integrals, finding
\begin{equation}
S_\lambda(\tau)= \tau\frac{e^4 n_1(n_0+1) \Omega V}{2 \epsilon_r^2 c^3 MN \omega_0}\frac{\omega_2}{\omega_1}  \left |R_{ijl}  (e_1^i)^* e_2^j e^l_{\lambda} \right |^2,
\end{equation}
where $\Omega$ is the solid angle spanned by the detector and $c$ is the speed of light. Second, we divide the scattering probability by the number of incident photons $N_1$ hitting an area $A=L_x L_y$ of the sample during a time $\tau$, i.e. $N_1=c A \tau/V =c \tau /L_z$. We find the same Raman efficiency calculated in the Ref.~\cite{loudon1963},
\begin{equation}
I_\lambda= \frac{e^4 n_1(n_0+1) \Omega V L_z}{2 \epsilon_r^2 c^4 MN \omega_0}\frac{\omega_2}{\omega_1} \left |R_{ijl}  (e_1^i)^* e_2^j e^l_{\lambda} \right |^2\equiv \gamma \frac{\omega_2}{\omega_1}  \left |{\bf e}_1^\dagger \cdot {\bf R}_{\lambda}  \cdot {\bf e}_2  \right |^2,
\end{equation}
which is the expression used in Eq. (\ref{eq:rint}). In the last equality of the preceding expression, we have used $(R_\lambda)_{i j} \equiv R_{i j l} e^l_\lambda$.
The factor $\gamma$ is approximately independent of the frequency of the incident light. Likewise, the factor $\omega_2/\omega_1$ is approximately equal to one. We have made these simplifying assumptions in the main text when plotting $I_\lambda$ as a function of $\omega_1$.

We note that, upon summing over phonon modes $\lambda$, upon dividing by the solid angle $\Omega$, and upon taking $n_1=1$ and $n_0=0$, our expression of $I_\lambda$ matches Eq. (7) of Ref.~\cite{Basko:2009}. The equivalence is realized via the correspondence
\begin{equation}
\frac{\omega_2}{\pi c^2} {\cal M}^{i j l} \leftrightarrow \sqrt{\gamma\frac{\omega_2}{\omega_1}} R_{i j l},
\end{equation}
where ${\cal M}$ is defined in Eq. (8) of Ref.~\cite{Basko:2009}.

 \section{A symmetry-induced constraint on the Raman tensors of in-plane phonons}
\label{sec:sym}

As mentioned in the main text, the 2D tight-binding model on a honeycomb lattice with Semenoff and Haldane mass terms has $C_{3h}$ symmetry.  For in-plane degenerate phonons of $C_{3h}$ group, symmetry predicts the following Raman tensors \cite{Cardona1982}: 
 \begin{align}
 \label{eq:C3h_in}
   R_{E_2(x)} &= \begin{pmatrix}
   R_{xx}^{\rm ip} & R_{xy}^{\rm ip}\\
   R_{xy}^{\rm ip} & -R_{xx}^{\rm ip}
    \end{pmatrix}\notag\\
       R_{E_2(y)} &= \begin{pmatrix}
    R_{xy}^{\rm ip}  & -R_{xx}^{\rm ip} \\
    -R_{xx}^{\rm ip} & -R_{xy}^{\rm ip}
    \end{pmatrix}.
\end{align}
Contrary to the case of the out-of-plane phonon, the Raman tensors for in-plane phonons are symmetric and they realize the inequality $|R_{xy}^{\rm ip}|\ll |R_{xx}^{\rm ip}|$ for all $\omega_1\gg \omega_0$. 
The purpose of this appendix is to explain the second property by showing that the coefficient $R_{xy}^{\rm ip}$ in Eq. (\ref{eq:C3h_in}) vanishes for in-plane phonons when the phonon frequency is neglected. 

In the $C_{3h}$ group, the ${\cal M}_y$ mirror is broken by the Haldane mass term of the electronic Hamiltonian. However, ${\cal M}_y$ mirror followed by  time reversal symmetry ${\cal T}$ is still a symmetry of the system \cite{mirrorsymmetry}:
\begin{equation}
\label{eq:mt}
h_{\rm e}({\bf k}') = {\cal S}^{-1} h_{\rm e}({\bf k}) {\cal S},
\end{equation}
where ${\cal S}={\cal M}_y {\cal T}$ and ${\bf k}'\equiv {\cal S} {\bf k} =(-k_x, k_y)$. Also, ${\cal S}^{-1} h_{\rm e}({\bf k}) {\cal S}=h_{\rm e}({\bf k})^*$, as both ${\cal T}$ and ${\cal M}_y$ acts as identity matrices in the $\{A, B\}$ sublattice space.
Equation (\ref{eq:mt}) immediately implies 
\begin{equation}
\label{eq:ensym}
\epsilon_{{\bf k} m} = \epsilon_{{\bf k}' m}
\end{equation}
for the energies of a band $m$, and 
\begin{equation}
{\cal S} |m, {\bf k}\rangle = |m, {\bf k}'\rangle
\end{equation}
for the band eigenstates (modulo an arbitrary and herein unimportant phase factor).
Antiunitarity of ${\cal S}$ imposes
\begin{equation}
\langle m { \bf k}'| m^\prime {\bf k}' \rangle = \langle m { \bf k}| m' {\bf k} \rangle^*
 \end{equation}
 and 
 \begin{equation}
 \langle m { \bf k}| O({\bf k})|m^\prime {\bf k} \rangle = \langle m {\bf k}'| {\cal S} O({\bf k}) {\cal S}^{-1}|m^\prime {\bf k}' \rangle^*
 \end{equation}
 for any $k-$dependent operator $O$. Using this relation, we can verify the following relations for the matrix elements of the electron-phonon interaction for in-plane phonons:
 \begin{align}
 &\frac{1}{a}\frac{\partial t_1}{\partial a}\left({\bf u}_A-{\bf u}_B\right)_\alpha  \left \langle m {\bf k} \left | \sum_i {a}_{\alpha i}\left[\cos({\bf k}\cdot{\bf a}_i)\sigma_x+\sin({\bf k}\cdot{\bf a}_i)\sigma_y\right]  \right |m^\prime {\bf k} \right \rangle \notag  \\ 
 =& (-1)^{\delta_{\alpha y}}\frac{1}{a}\frac{\partial t_1}{\partial a}\left({\bf u}_A-{\bf u}_B\right)_\alpha  \left \langle m {\bf k^\prime} \left | \sum_i {a}_{\alpha i}\left[\cos({\bf k^\prime}\cdot{\bf a}_i)\sigma_x+\sin({\bf k^\prime}\cdot{\bf a}_i)\sigma_y\right]  \right |m^\prime {\bf k^\prime} \right \rangle^*,
\end{align}
where $\alpha\in\{x, y\}$.
Concerning the matrix elements of the electron-photon interaction, we have
 \begin{align}
  -e { A}_\alpha \left \langle m {\bf k} \left |\frac{\partial h_{\rm e}({\bf k})}{\partial { k}_\alpha} \right |m^\prime {\bf k} \right \rangle &= -(-1)^{\delta_{\alpha x}}e { A}_\alpha \left \langle m {\bf k^\prime} \left |\frac{\partial h_{\rm e}({\bf k^\prime})}{\partial  k^\prime_\alpha} \right |m^\prime {\bf k^\prime} \right \rangle^*. \\
  \frac{e^2}{2}  A_\alpha A_\beta  \left \langle m {\bf k} \left |  \frac{\partial^2 h_{\rm e}({\bf k})}{\partial k_\alpha \partial k_\beta} \right | m^\prime {\bf k} \right \rangle &= (-1)^{\delta_{\alpha y}+ \delta_{\beta y}}\frac{e^2}{2}  A_\alpha A_\beta  \left \langle m {\bf k^\prime} \left |  \frac{\partial^2 h_{\rm e}({\bf k^\prime})}{\partial k^\prime_\alpha \partial k^\prime_\beta} \right | m^\prime {\bf k^\prime} \right \rangle^*.
\end{align}
Finally, for the coupled electron-phonon electron-photon interaction, we get
 \begin{align}
 -e { A_\alpha} \left \langle m {\bf k} \left |\frac{\partial h^\gamma_{\rm e-pn, in}({\bf k})}{\partial {k_\alpha}}\right |m^\prime {\bf k} \right \rangle &=-(-1)^{\delta_{\gamma x}+\delta_{\alpha y}}e {A_{\alpha}} \left \langle m {\bf k}^\prime \left | \frac{\partial  h^\gamma_{\rm e-pn, in}({\bf k^\prime})}{\partial {k^\prime_{\alpha}}} \right |m^\prime {\bf k}^\prime \right \rangle^*.  \\
\frac{e^2}{2} A_\alpha A_\beta \left \langle m {\bf k} \left|\frac{\partial^2 h^\gamma_{\rm e-pn, in}({\bf k})}{\partial k_\alpha \partial k_\beta} \right | m^\prime {\bf k}\right \rangle   &=\frac{e^2}{2} A_\alpha A_\beta (-1)^{\delta_{\alpha x}+\delta_{\beta x}+\delta_{\gamma y}}\left \langle m {\bf k}^\prime \left|\frac{\partial^2 h^{\gamma}_{\rm e-pn, in}({\bf k}^\prime)}{\partial k_\alpha^\prime \partial k_\beta^\prime} \right | m^\prime {\bf k}^\prime \right \rangle^*.
\end{align}
Here, $\gamma=x$ for $E_2(x)$ phonons and $\gamma=y$ for $E_2(y)$ phonons. 

Substituting the preceding matrix elements and Eq. (\ref{eq:ensym}) in the microscopic expression for the Raman tensor, and then redefining ${\bf k}'\to {\bf k}$ by a change of variables in the electronic wave vector sum contained in $R_{i j}$, we arrive at the relation
 \begin{align}
 \label{eq:Rijrel1}
  R_{ij}(-\omega_1,\omega_2,\omega_0;\eta) =\left\{\begin{array}{c} (-1)^{1+\delta_{ij}} R_{ij}^*(-\omega_1,\omega_2,\omega_0;-\eta) \text{    , for $E_2(x)$ phonons}\\
  (-1)^{\delta_{ij}} R_{ij}^*(-\omega_1,\omega_2,\omega_0;-\eta) \text{    , for $E_2(y)$ phonons,}\end{array}\right.
  \end{align}
 where we have used essentially the same notation as in Refs.~\cite{loudon1963,tensorrelation} (except for the fact that here we employ $\eta$  instead of $\Gamma$ to denote the electronic decay rate). 
 
However, the Raman tensor elements obey the general relations \cite{tensorrelation}
  \begin{align}
\label{eq:Rijrel2}
  R_{ij}(-\omega_1,\omega_2,\omega_0;\eta) &= R_{ij}^*(\omega_1,-\omega_2,-\omega_0;-\eta)\notag\\
  R_{ij}(-\omega_1,\omega_2,\omega_0;\eta) &= R_{ji}(\omega_2,-\omega_1,\omega_0;\eta).
 \end{align}
 Strictly speaking, Ref. \cite{tensorrelation} found Eq. (\ref{eq:Rijrel2}) for the Raman amplitude treated in Ref. \cite{loudon1963}, but the equation applies also for the additional amplitudes included in Ref. \cite{Basko:2009} and in our present work. 
  
 Combining Eqs. (\ref{eq:Rijrel1}) and (\ref{eq:Rijrel2}), we arrive at
\begin{align}
 \label{eq:Rijrel3}
  R_{ij}(-\omega_1,\omega_2,\omega_0;\eta) =\left\{\begin{array}{c} (-1)^{1+\delta_{ij}} R_{ji}(-\omega_2,\omega_1,-\omega_0;\eta) \text{    , for $E_2(x)$ phonons}\\
  (-1)^{\delta_{ij}} R_{ji}(-\omega_2,\omega_1,-\omega_0;\eta)  \text{    , for $E_2(y)$ phonons.}\end{array}\right.
  \end{align}
  Now, in the circumstance in which the phonon frequency is negligible (an approximation that is often adopted since the phonon frequency is commonly much smaller than the incoming light frequency), we set $\omega_0\to 0$ and $\omega_2\to \omega_1$ in Eq. (\ref{eq:Rijrel3}), so that 
\begin{align}
 \label{eq:Rijrel4}
  R_{ij}(-\omega_1,\omega_1,0;\eta) =\left\{\begin{array}{c} (-1)^{1+\delta_{ij}} R_{ji}(-\omega_1,\omega_1,0;\eta) \text{    , for $E_2(x)$ phonons}\\
  (-1)^{\delta_{ij}} R_{ji}(-\omega_1,\omega_1,0;\eta)  \text{    , for $E_2(y)$ phonons.}\end{array}\right.
  \end{align}
 Two immediate consequences follow from Eq. (\ref{eq:Rijrel4}).
 First, $R_{xy}=-R_{yx}$ for the $E_2(x)$ phonon. 
 But $C_{3h}$ symmetry imposes $R_{xy}=R_{xy}$ [cf. Eq. (\ref{eq:C3h_in})]. 
To reconcile these two results, the coefficient $R_{xy}^{\rm ip}$ of Eq. (\ref{eq:C3h_in}) must vanish when the phonon frequency is neglected. 
 The second consequence of Eq. (\ref{eq:Rijrel4}) is that, for $E_2(y)$ phonons, $R_{ii}=-R_{ii}$.
 Consequently, once again the coefficient $R_{xy}^{\rm ip}$ of Eq. (\ref{eq:C3h_in}) must vanish when the phonon frequency is neglected. 
  
 To summarize, we find that the ${\cal M}_y {\cal T}$ symmetry of the hexagonal tight-binding model with Haldane and Semenoff masses imposes that the Raman tensors for the in-plane phonons have the form
  \begin{align}
   R_{E_2(x)} &\simeq \begin{pmatrix}
   R_{xx}^{\rm ip} & 0 \\
    0& -R_{xx}^{\rm ip}
    \end{pmatrix}\notag\\
       R_{E_2(y)} &\simeq  \begin{pmatrix}
    0  & -R_{xx}^{\rm ip}\\
    -R_{xx}^{\rm ip} & 0
    \end{pmatrix}
\end{align}
when the phonon frequency is neglected. If the small but nonzero phonon frequency is taken into account, then the correct form of the Raman tensor is given by Eq. (\ref{eq:C3h_in}), with $|R_{xy}^{\rm ip}|\ll|R_{xx}^{\rm ip}|$. We have verified this statement in our numerical calculation of the Raman tensor. \par 
\end{widetext}
\bibliography{referenceFile}

\end{document}